\documentclass[11pt]{article}
\usepackage{jheppub}
\usepackage{amsmath}
\usepackage{amssymb}
\usepackage{braket}
\usepackage{bbold}
\usepackage{pifont}
\usepackage{tensind}
\usepackage{color}
\usepackage{cancel}
\usepackage{tikz}
\usetikzlibrary{trees,decorations.pathmorphing,decorations.markings}
\tensordelimiter{`}

\newcommand{\gmn}{g_{\mu\nu}}
\newcommand{\fmn}{f_{\mu\nu}}
\newcommand{\hmn}{h_{\mu\nu}}
\newcommand{\dG}{\delta G}
\newcommand{\dM}{\delta M}
\newcommand{\beqn}{\begin{eqnarray}}
\newcommand{\eeqn}{\end{eqnarray}}
\newcommand{\dd}{\mathrm{d}}
\newcommand{\nn}{\nonumber}

\newcommand{\bgmn}{\bar{g}_{\mu\nu}}

\newcommand{\pa}{\partial}
\newcommand{\mpl}{m_\mathrm{Pl}}
\newcommand{\be}{\begin{equation}}
\newcommand{\ee}{\end{equation}}

\newcommand{\ph}{\phantom}
\newcommand{\td}{\mathrm{d}}

\tikzset{
    graviton/.style={decorate, decoration={coil,amplitude=4pt, segment length=5pt}, semithick, double, draw=red},
    Mgraviton/.style={decorate, decoration={snake,amplitude=5pt}, semithick, double, draw=magenta},
    sm/.style={draw=blue,double,thick},
}

\title{Dark matter scenarios with multiple spin-2 fields}

\author[1]{N.L. Gonz\'alez Albornoz,}
\author[1]{Angnis~Schmidt-May,}
\author[2]{Mikael von Strauss}
\affiliation[1]{Arnold Sommerfeld Center for Theoretical Physics,\\ 
Ludwig-Maximilians-Universit\"at M\"unchen, Theresienstra\ss e 37, D-80333 Munich, Germany}
\affiliation[2]{Nordita, KTH Royal Institute of Technology and Stockholm University, 
Roslagstullsbacken 23, SE-106 91 Stockholm, Sweden}

\emailAdd{Nicolas.Gonzalez@physik.uni-muenchen.de}
\emailAdd{A.SchmidtMay@physik.uni-muenchen.de}
\emailAdd{Mikael.von.Strauss@su.se}

\abstract{
We study ghost-free multimetric theories for $(N+1)$ tensor fields with a coupling to matter 
and maximal global symmetry group $S_N\times(Z_2)^N$. Their mass spectra contain a massless mode, 
the graviton, and $N$ massive spin-2 modes. One of the massive modes is distinct by being the heaviest, 
the remaining $(N-1)$ massive modes are simply identical copies of each other. 
All relevant physics can therefore be understood from the case $N=2$.
Focussing on this case, we compute the full perturbative action up to cubic order and derive 
several features that hold to all orders in perturbation theory.  
The lighter massive mode does not couple to matter and neither of the massive modes 
decay into massless gravitons. 
We propose the lighter massive particle as a candidate for dark matter
and investigate its phenomenology in the parameter region where the matter coupling
is dominated by the massless graviton. The relic density of massive spin-2 can originate
from a freeze-in mechanism or from gravitational particle production, giving rise to two 
different dark matter scenarios. 
The allowed parameter regions are very different from those in scenarios with only one massive spin-2 field
and more accessible to experiments.
}


\begin{document} 
\begin{flushright}
\vspace{10pt} \hfill{NORDITA 2017-095} \\
\hfill{LMU-ASC 56/17} \vspace{20mm}
\end{flushright}
\maketitle
\flushbottom

%%%%%%%%%%%%%%%%%%%%%%%%%%%%%%%%%%%%%%%%%%%%%%%%%%%%%%%%%%%
\section{Introduction}
%%%%%%%%%%%%%%%%%%%%%%%%%%%%%%%%%%%%%%%%%%%%%%%%%%%%%%%%%%%

Field theories with massive spin-2 degrees of freedoms have developed into a topic of interest
in particle physics and cosmology; see~\cite{Hinterbichler:2011tt, deRham:2014zqa, Schmidt-May:2015vnx}
for recent reviews. Massive spin-2 particles may be viewed as a natural
addition to the Standard Model (SM) combined with General Relativity (GR), since the first 
contains massive and massless particles up to spin-1 and the latter describes a massless 
spin-2 particle, the graviton. The linear Fierz-Pauli theory for massive spin-2 
was already developed in 1939~\cite{Fierz:1939ix}, but it suffers from
a discontinuous zero-mass limit, known as the vDVZ discontinuity~\cite{vanDam:1970vg, Zakharov:1970cc}. 
Due to this, the Fierz-Pauli theory is not compatible with basic solar system tests. A possible
solution to this problem is the inclusion of nonlinear interactions giving rise to a Vainshtein
screening mechanism~\cite{Vainshtein:1972sx} that can cure the discontinuity. 
Unfortunately, for decades nonlinear interactions for massive spin-2 fields
were believed to generally contain a fatal ghost instability~\cite{Boulware:1973my}. 
This problem was resolved only a few years ago when a set of consistent interactions
was identified and proven to avoid the ghost 
\cite{deRham:2010ik, deRham:2010kj, Hassan:2011vm, Hassan:2011hr, Hassan:2011tf, Hassan:2011ea, Hassan:2012qv}.
The resulting theory contains a massive tensor field $\gmn$ 
that can couple to SM matter and thereby mediate gravitational 
interactions. Since the gravitational force is long-ranged, 
the spin-2 mass is constrained to be extremely small (essentially on the order of the Hubble scale) 
in this setup. 

The ghost-free massive gravity action also contains a second tensor field
$\fmn$ which is non-dynamical and merely acts as a fixed reference metric. 
It has later been realised that, in perturbation theory around any background solution 
$\gmn$, the reference metric can be traded against certain functions of the curvature 
of $\gmn$.~\cite{Hassan:2013pca, Bernard:2014bfa, Bernard:2015mkk, Bernard:2015uic, Mazuet:2017hey}. 
Completely new possibilities opened when it was shown that the second tensor can be given its own
dynamics without reintroducing the ghost~\cite{Hassan:2011zd, Hassan:2011ea}. This results
in a bimetric theory for gravity, describing nonlinear interactions of massless and massive spin-2 fields
and their couplings to SM matter~\cite{Hassan:2012wr}.

Due to the presence of the massless mode
that can mediate a long-range force, the constraints on the spin-2 mass are now much less severe. 
In fact, it has been shown that the gravitational interactions in bimetric theory can resemble 
those of GR to any precision for any value of the mass
\cite{Baccetti:2012bk, Hassan:2014vja, Akrami:2015qga, Babichev:2016bxi}.
This is achieved by weakening the coupling of the massive spin-2 field to the matter sector. 
Interestingly, this does not affect the couplings between massive and massless spin-2 fields
and the massive mode continues to gravitate with the same strength as the SM fields.

These insights suggest that a relic density of massive spin-2 particles could act as a 
dark matter (DM) component in the Universe. The existence of DM has been inferred from various
astrophysical and cosmological observations but the particle has never been seen other than through
its gravitational interactions. In standard scenarios, a model-dependent production mechanism is 
responsible for creating the DM relic density in the early Universe.\footnote{Other examples of 
models for DM which do not assume its particle nature 
include modifications of the gravitational laws~\cite{Milgrom:2014usa} 
and primordial black holes~\cite{Carr:2016drx}.}
Additional symmetries or the very weak interactions of the DM 
particle ensure its stability until the present time. 
So far, all attempts to produce or detect the DM particle have remained unsuccessful 
(see e.g.~\cite{Agashe:2014kda, Ackermann:2015lka}), suggesting that it is very heavy and/or
its interactions with baryonic matter are simply too weak.

Pursuing (rather speculative) ideas first mentioned in~\cite{Maeda:2013bha, Schmidt-May:2016hsx},
a possible scenario for massive spin-2 DM was proposed and thoroughly 
analysed in~\cite{Babichev:2016hir, Babichev:2016bxi} (see also \cite{Aoki:2016zgp}). 
It was found that ghost-free bimetric theory provides a natural framework for a 
DM candidate with spin-2 and a mass of a few TeV. Its observed abundance can be explained by
invoking a non-thermal (``freeze-in") production mechanism and the decay products of the particle could 
in principle be seen in indirect detection experiments. For a recent review of the ``freeze-in" mechanism see~\cite{Bernal:2017kxu}. The authors of~\cite{Chu:2017msm} 
pointed out that strong spin-2 self-interactions can affect the production mechanism by 
allowing for a thermalisation of the dark sector, and the DM mass could be as low as 1\,MeV. 
Ideas related to these scenarios and extensions were developed 
in~\cite{Marzola:2017lbt, Aoki:2017ffl, Aoki:2017cnz}. 
Unfortunately, while providing a theoretically well-motivated model, bimetric theory predicts
a DM candidate whose interactions with baryonic matter are too tiny to ever be directly detected. 
For the same reason, it is impossible to produce the massive spin-2 field in colliders. 
Unless one is lucky and observes its decay products in indirect detection experiments, there
are few possibilities to test the model.\footnote{There is a chance that the spin-2 self-interactions
give rise to observable effects in DM halos, see~\cite{Chu:2017msm}.} 
 
It is an interesting question how the phenomenology of massive spin-2 fields 
is affected by considering more than one species of them.
The corresponding ghost-free multimetric interactions were first considered in the vierbein language 
in~\cite{Hinterbichler:2012cn}. The vierbeins are subject to certain 
symmetrisation constraints which ensure the absence of ghost 
instabilities~\cite{Deffayet:2012zc, deRham:2015cha}
and at the same time allow the interactions to be expressed in terms 
of metric tensors alone~\cite{Hinterbichler:2012cn, Hassan:2012wt}.
The resulting set of consistent theories contain pairwise bimetric interactions between the
tensor fields which are not allowed to form a closed loop~\cite{Nomura:2012xr}.
They describe one massless spin-2 interacting with several massive spin-2 fields.
Their mass spectra were investigated in~\cite{Baldacchino:2016jsz}; 
a first analysis of cosmological solutions for theories with three tensors was 
performed in~\cite{Luben:2016lku}. 
For more related work, see~\cite{Noller:2013yja, Noller:2015eda, Scargill:2015wxs}.

Bimetric theory can be viewed as the extension of GR by one massive tensor and na\"ively one
may expect that adding more tensor fields will not lead to new physical effects. 
The aim of this paper is to demonstrate the opposite and present some of the immediate 
implications for DM phenomenology.

\paragraph{Summary of results.}

\begin{itemize}

\item
An underlying reason for the occurrence of new structures in multimetric 
interactions with respect to bimetric theory is the possibility to make 
them invariant under discrete symmetry groups.
We argue that the maximal global symmetry in multimetric theories with $(N+1)$ tensors 
and a coupling to matter is $S_N\times(Z_2)^N$ and we identify the corresponding actions.

\item
The mass spectrum of models with $S_N\times(Z_2)^N$ invariance is  
used to study features of the perturbative action.
Moreover, we explicitly compute all cubic interaction vertices in terms of the mass
eigenstates for the case $N=2$ which we refer to as ``trimetric theory".

\item We find that trimetric theory with maximal global symmetry is a nontrivial generalisation 
of the $N=1$ case. Just like bimetric theory, it contains a parameter $\alpha$ which 
together with the spin-2 mass scale controls the deviations from GR. 
We focus on the case $\alpha < 1$ for which these deviations are
small over a large mass range. 
The heaviest spin-2 field behaves very similarly to the massive mode 
of bimetric theory; it does not decay into massless gravitons and
not into lighter spin-2 fields.
Its couplings to SM matter are extremely weak.
The lighter massive field, however, possesses new features. In particular, it does 
not interact with matter at all and it cannot decay into other spin-2 particles either. 
Hence, it is entirely stable. 
No new phenomena occur for more than three fields; the additional modes all have 
equal mass and are simply identical copies of the lighter mode in trimetric theory.

\item We propose that the observed DM abundance could be made up from the massive spin-2 
particles of maximally symmetric trimetric theory. For $\alpha \lesssim 10^{-12}$, this scenario 
essentially gives rise to a phenomenology that is indistinguishable from the bimetric one. 
However, in the parameter region where the heaviest mode is not stable enough to contribute to the DM abundance, 
indirect detection experiments and stability of DM do not put an upper bound on the parameter 
$\alpha$. In this case, the lighter spin-2 mode is the only DM candidate, giving rise to an entirely
new phenomenology.
Its relic density can be produced either through a freeze-in mechanism or
via gravitational particle production. In the first scenario, the bounds on the spin-2 mass $m$ are,
\beqn
1\,\mathrm{TeV}\lesssim m \lesssim 10^{11}\,\mathrm{GeV}\,,
\eeqn
whereas the second production mechanism requires 
\beqn
m \gtrsim 10^{10}\,\mathrm{GeV}\,.
\eeqn

\item In the above parameter region with $\alpha<1$, 
the massive spin-2 particles unfortunately still remain unobservable,
at least through direct detection experiments or production in colliders. 
However, our results suggest that, in contrast to the bimetric DM scenario, 
the trimetric model contains a new interesting region in its parameter space,
corresponding to $\alpha>1$. Here it may become possible to test the model 
with astrophysical and cosmological observations or even detect 
the heaviest spin-2 particle directly.

\end{itemize}

\paragraph{Organisation of the paper.}
The paper is organised as follows.
In section~\ref{sec:mult} we briefly review multimetric interactions before identifying the
theories with maximal global symmetries.
The mass spectrum of maximally symmetric trimetric theory 
is reviewed in section~\ref{sec:trim}.
Section~\ref{sec:perturbact} discusses the perturbative expansion of the trimetric
action and the generalisation to multiple fields.
The implications on DM phenomenology are studied in section~\ref{sec:dm}.
We discuss our results and give an outlook in section~\ref{sec:discussion}.
Appendix~\ref{app:cubic} contains the full expressions for cubic vertices in trimetric theory.

%%%%%%%%%%%%%%%%%%%%%%%%%%%%%%%%%%%%%%%%%%%%%%%%%%%%%%%%%%%
\section{Multimetric theory}\label{sec:mult}
%%%%%%%%%%%%%%%%%%%%%%%%%%%%%%%%%%%%%%%%%%%%%%%%%%%%%%%%%%%
We begin by defining the structure of multimetric actions
and discuss how imposing global symmetries singles out a class
of models with a significantly smaller parameter space.

%%%%%%%%%%%%%%%%%%%%%%%%%%%%%%%%%%%%%%%%%%%%%%%%%%%%%%%%%%%
\subsection{Multimetric interactions}
%%%%%%%%%%%%%%%%%%%%%%%%%%%%%%%%%%%%%%%%%%%%%%%%%%%%%%%%%%%

We consider a set of $(N+1)$ tensor fields $\fmn^{(i)}$, $i=1,\hdots, N+1$ 
and their ghost-free couplings. The fields can only interact pairwise 
with each other and the interaction graphs cannot contain closed loops.
The pairwise interactions arise through a potential,
\begin{align}\label{intform}
S_\mathrm{int}[f^{(i)},f^{(j)}]=-2m^2\int\dd^4x~\sqrt{|g|}\,V(f^{(i)},f^{(j)};\beta^{(ij)}_n)\,,
\end{align}
whose structure is almost entirely fixed by requiring the absence of ghosts.\footnote{For the 
long list of references containing the steps of the consistency proof, we refer the reader to
the introduction.} 
The parameter $m^2$ in front has mass dimension two 
and the potential can be written in the form,
\beqn\label{intpot}
V(g,f;\beta_n)=\sum_{n=0}^4\beta_n e_n\big(\sqrt{g^{-1}f}\,\big)\,,
\eeqn
with a set of dimensionless parameters $\beta_n$, $n=0,\hdots 4$.
The (1,1) tensor $\big(\sqrt{g^{-1}f}\,\big)^{\mu}_{~\nu}$ is a matrix square-root defined via 
$\big(\sqrt{g^{-1}f}\,\big)^2=g^{-1}f$.\footnote{For a careful treatment of the
square-root matrix and its precise meaning in bimetric theory, see~\cite{Hassan:2017ugh}.}  
The elementary symmetric polynomials $e_n(S)$ are scalar functions of the matrix argument $S$
and can be defined recursively through,
\beqn
e_n(S)=\frac{(-1)^{n+1}}{n}\sum_{k=0}^{n-1}(-1)^k\mathrm{Tr}(S^{n-k})e_k(S)\,,\qquad e_0(S)=1\,.
\eeqn
We furthermore introduce a coupling to SM matter fields 
which we collectively denote by $\phi$.
This coupling must have the same form as in GR,
\begin{align}
S_\mathrm{matter}[g,\phi]=\int d^4x~\sqrt{|g|}\,\mathcal{L}_\mathrm{matter}(g,\phi).
\end{align}
It is not possible to couple more than one of the metrics to matter without
reintroducing ghosts~\cite{Yamashita:2014fga, deRham:2014naa}.
The coupling thus picks out a ``physical metric", which we shall denote by $\gmn$.
The remaining fields, which do not directly interact with matter, will be labelled $\fmn^{(i)}$ 
where now $i=1,\hdots, N$. 

   \begin{figure}[h]
    \begin{center}
    \includegraphics[width=170pt]{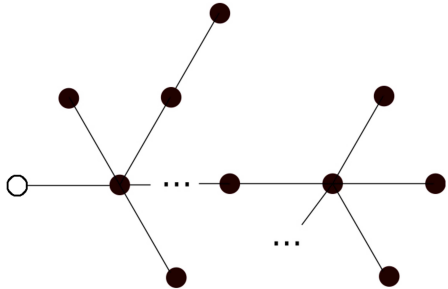}
    \caption{Theory graph of a general ghost-free multimetric theory. 
    The black circles represent the $(N+1)$ metrics, 
    the black lines represent pairwise interactions
    and the white circle represents the matter sector.}
    \end{center}
    \end{figure}

Clearly, the above restrictions still leave us with some freedom in constructing
a multimetric model with $(N+1)$ fields. 
The case $N=0$ corresponds to GR with metric $\gmn$. 
There is a unique way to add a second tensor $\fmn$: it is coupled to $\gmn$ through an 
interaction of the form (\ref{intform}), resulting in the well-studied ghost-free
bimetric theory. A third tensor field can now be added in two different ways: Either 
through a coupling to $\gmn$ or to $\fmn$. For $N>3$ fields, one can construct 
interaction diagrams of increasing complexity. 
This is illustrated in Figure 1. The only restriction on these graphs is that the lines
representing pairwise interactions of the form (\ref{intform}) may never form a closed loop.

%%%%%%%%%%%%%%%%%%%%%%%%%%%%%%%%%%%%%%%%%%%%%%%%%%%%%%%%%%%%%%%%%%
\subsection{Models with maximal global symmetry}\label{sec:mwms}
%%%%%%%%%%%%%%%%%%%%%%%%%%%%%%%%%%%%%%%%%%%%%%%%%%%%%%%%%%%%%%%%%%

Due to the large number of possibilities for different multimetric theories, it would be
useful to have a principle that singles out a certain class of preferred models. 
Here we demand the presence of global symmetries in the multimetric action. 
This will result in a simple class of multimetric models with a 
significantly smaller number of interaction parameters.

%%%%%%%%%%%%%%%%%%%%%%%%%%%%%%%%%%%%%%%%%%%%%%%%%%%%%%%%%%%%%%
\subsubsection{Interchange symmetry $S_N$}\label{sec:intsym}
%%%%%%%%%%%%%%%%%%%%%%%%%%%%%%%%%%%%%%%%%%%%%%%%%%%%%%%%%%%%%%

We are looking for multimetric theories with the maximum amount of discrete, global symmetries.
One obvious symmetry of this type corresponds to the interchange of fields in a theory graph.
Since the matter coupling necessarily breaks part of that symmetry by singling out a 
physical metric, the best one can achieve is a graph that is symmetric under the permutation
of the remaining $N$ fields.
The graph corresponding to a theory for which such a symmetry is possible is displayed in Figure~2.
  \begin{figure}[h]
    \begin{center}
    \includegraphics[width=90pt]{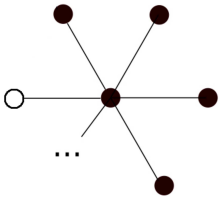}
    \caption{Theory graph of a multimetric theory with interchange symmetry
    among the $N$ satellite fields. }
    \end{center}
    \end{figure}
The action associated to this graph is of the form,
\begin{align}\label{multi}
S[g,f^{(i)}]=m_g^2\int d^4x\Bigg(\sqrt{|g|}&\,R(g)+\sum_{i=1}^N\alpha_i^2\sqrt{|f^{(i)}|}\,R(f^{(i)})
\nn\\
&-2m^2\sum_{i=1}^N\sqrt{|g|}\,V(g,f^{(i)};\beta^{(i)}_n)\Bigg)+S_\mathrm{matter}[g,\phi]\,.
\end{align}
The first line contains the Einstein-Hilbert kinetic terms for the tensor fields 
with ``Planck masses" $m_g$ and $\alpha_im_g$. The second line parameterises the 
interactions.

Requiring the invariance of the theory under the $S_N$ symmetry,
\beqn
\fmn^{i} \longleftrightarrow \fmn^{j}\,, \qquad \forall~~i,j = 1,\hdots, N\,,
\eeqn
restricts the interaction parameters in (\ref{multi}). The conditions on them can be derived as follows.
After rescaling the satellite fields, 
$\fmn^{(i)}\rightarrow\tilde{f}_{\mu\nu}^{(i)}=\alpha_i^{-2}\fmn^{(i)}$,
the Einstein-Hilbert terms are symmetric under interchange of $\tilde{f}_{\mu\nu}^{(i)}$
and $\tilde{f}_{\mu\nu}^{(j)}$.
Since $e_n(\lambda S)=\lambda^ne_n(S)$ for a scalar $\lambda$, 
the potential is invariant only if one imposes,
\beqn\label{symmcond}
\alpha^{-n}_i\beta^{(i)}_n=\alpha^{-n}_j\beta^{(j)}_n
\qquad
\forall~~ i,j=1,\hdots,N\,.
\eeqn
These conditions imply that we are left with only five free interaction parameters, since we can now 
write $\beta^{(i)}_n=\alpha_i^n\beta_n$ for all $i$ with some fixed set of parameters $\beta_n$.

One may also consider slightly less symmetric examples where the interchange symmetry is 
only realised among $k$ out of the $N$ fields. This is the case when $\beta^{(i)}_n=\alpha_i^n\beta_n$
for $i=1,\hdots,k$ while the remaining $\beta^{(j)}_n$ remain arbitrary. Obviously, the discrete symmetry
group in this case is $S_k$.

%%%%%%%%%%%%%%%%%%%%%%%%%%%%%%%%%%%%%%%%%%%%%%%%%%%%%%%%%%%%%%%
\subsubsection{Reflection symmetry $(Z_2)^N$}\label{sec:reflsym}
%%%%%%%%%%%%%%%%%%%%%%%%%%%%%%%%%%%%%%%%%%%%%%%%%%%%%%%%%%%%%%%

The maybe less obvious set of symmetries that one can demand in addition to the one in the previous 
subsection are the transformations,
\beqn\label{reflsym}
\sqrt{g^{-1}f^{(i)}}\longrightarrow - \sqrt{g^{-1}f^{(i)}}\qquad \forall~~ i=1,\hdots,N\,.
\eeqn
Given a square root $S$ as a solution to the equation $(S^2)^\mu_{~\nu}=g^{\mu\rho}f_{\rho\nu}$, 
its negative counterpart $-S$ will also be a solution to the same equation. 
One way to implement the above transformation is to express the
square-root matrix in terms of two constrained vierbeins\footnote{A multi-spin-2 action in terms
of vierbeins is only equivalent to a multimetric action when one imposes certain symmetrisation
constraints that allow for an explicit evaluation of the square root~\cite{Hinterbichler:2012cn, Hassan:2012wt}.} 
$e^a_{~\mu}$ and $v^a_{~\mu}$ with $\gmn=\eta_{ab}e^a_{~\mu}e^b_{~\nu}$ 
and $\fmn=\eta_{ab}v^a_{~\mu}v^b_{~\nu}$. In terms of these we have,
\beqn
S^\mu_{~\nu}=\big(\sqrt{g^{-1}f}\,\big)^\mu_{~\nu}= e_b^{~\mu} v^b_{~\nu}\,.
\eeqn
Now for each vierbein $(v^{(i)})^b_{~\nu}$ of $\fmn^{(i)}$, we consider the transformations,
\beqn
(v^{(i)})^b_{~\nu}\longrightarrow -(v^{(i)})^b_{~\nu}\,,
\qquad 
e^a_{~\mu}\longrightarrow e^a_{~\mu}\,,
\eeqn
which leave all metrics invariant but transforms the square roots according to (\ref{reflsym}).
The Einstein-Hilbert terms are invariant since they contain the vierbeins only through their 
respective metrics. In the pairwise interactions the elementary symmetric polynomials
transform as,
\beqn
e_n(S)\longrightarrow (-1)^n e_n(S)\,.
\eeqn
Hence the potential is invariant only if we demand the odd terms to vanish,
\beqn\label{reflcond}
\beta^{(i)}_1=\beta^{(i)}_3=0\qquad \forall~~ i=1,\hdots,N\,.
\eeqn
It is again possible to impose a subset of symmetries
by demanding (\ref{reflcond}) to hold for only $k$ of the $N$ 
parameter sets. In this case the action will be invariant under
only $k$ transformations of the type (\ref{reflsym}) and the symmetry is $(Z_2)^k$.

%%%%%%%%%%%%%%%%%%%%%%%%%%%%%%%%%%%%%%%%%%%%%%%%%%%%%%%%%%%%%%%
\subsubsection{Maximally symmetric action}
%%%%%%%%%%%%%%%%%%%%%%%%%%%%%%%%%%%%%%%%%%%%%%%%%%%%%%%%%%%%%%%
Imposing the conditions (\ref{symmcond}) and (\ref{reflcond}) results in the multimetric 
actions with maximal amount of global symmetry, $S_N\times(Z_2)^N$. It reads,
\begin{align}\label{maxsym}
S[g,f^{(i)}]=m_g^2\int d^4x\Bigg(\sqrt{|g|}&\,R(g)+\sum_{i=1}^N\alpha_i^2\sqrt{|f^{(i)}|}\,R(f^{(i)})
\nn\\
&-2m^2\sum_{i=1}^N\sqrt{|g|}\,V(g,f^{(i)};\alpha_i^n\beta_n, )\Bigg)+S_\mathrm{matter}[g,\phi]\,,
\end{align}
where we have defined $\beta_n\equiv\alpha_i^{-n}\beta^{(i)}_n$ 
and the potentials now have the simple form,
\beqn\label{intpot2}
\sqrt{|g|}\,V(g,f^{(i)};\alpha_i^n\beta_n, )&=&
\beta_0\sqrt{|g|}\,+\alpha_i^2\beta_2
\sqrt{|g|}\,e_2\Big(\sqrt{g^{-1}f^{(i)}}\Big)+\alpha_i^4\beta_4\sqrt{|f^{(i)}|}\,\,,
\nn\\
e_2(S)&=&\frac1{2}\Big(S^\rho_{\ph\rho\rho}S^\sigma_{\ph\sigma\sigma}
-S^\rho_{\ph\rho\sigma}S^\sigma_{\ph\sigma\rho}\Big)\,.
\eeqn
We see that $\beta_0$ and $\beta_4$ simply parameterise cosmological constant
contributions\footnote{In fact, the $\beta_0$ terms in all potential contributions
are identical since $e_0(S)=1$ for any matrix $S$. 
We could have dropped $(N-1)$ of the degenerate parameters in the beginning but chose to include
them in order to shorten the expressions.}
and the only remaining interaction terms are proportional to $\beta_2$.
To our knowledge there exist no further global symmetries that can be imposed on 
the gravitational part of the action~(\ref{maxsym}).

%%%%%%%%%%%%%%%%%%%%%%%%%%%%%%%%%%%%%%%%%%%%%%%%%%%%%%%%%%%%%%%%%%%%%%%%
\section{Trimetric theory with maximal global symmetry}\label{sec:trim}
%%%%%%%%%%%%%%%%%%%%%%%%%%%%%%%%%%%%%%%%%%%%%%%%%%%%%%%%%%%%%%%%%%%%%%%%
In this section we work out the details of a multimetric theory with maximal global symmetry.
We specialise to three spin-2 fields to keep the expressions simple. The results obtained here
naturally generalise to $N$ fields, as will be discussed in section~\ref{sec:gen}.

The ghost-free action for three symmetric tensors $\gmn$, $\fmn$ and $\hmn$ 
with general parameters is of the form,
\begin{align}\label{triact}
\nn S[g,f,h]=m_g^2\int d^4x\Big(&\sqrt{|g|}\,R(g)+\alpha_f^2\sqrt{|f|}\,R(f)+\alpha_h^2\sqrt{|h|}\,R(h)\\
&-2m^2\sqrt{|g|}\,V(g,f;\beta_n^f)-2m^2\sqrt{|g|}\,V(g,h;\beta_n^h)\Big)+S_\mathrm{matter}[g,\phi]\,.
\end{align}
with potential as defined in (\ref{intpot}). We will start by discussing its mass spectrum and 
explaining the physical effects of imposing the global symmetry which in this case is
$S_2\times(Z_2)^2\simeq Z_2\times(Z_2)^2$.

%%%%%%%%%%%%%%%%%%%%%%%%%%%%%%%%%%%%%%%%%%%%%%%%%%%%%%%%%%%
\subsection{Proportional backgrounds}
%%%%%%%%%%%%%%%%%%%%%%%%%%%%%%%%%%%%%%%%%%%%%%%%%%%%%%%%%%%

In the following we review the mass spectrum of trimetric theory
which has been derived in~\cite{Baldacchino:2016jsz}, generalising the bimetric results of~\cite{Hassan:2012wr}.
We first need to find suitable maximally symmetric background solutions around which the fluctuations of the tensor fields
can be diagonalised into mass eigenstates. These are the proportional backgrounds 
obtained by solving the equations of motion with the ansatz,
\beqn
g_{\mu\nu}=c_f^{-2}f_{\mu\nu}=c_h^{-2}h_{\mu\nu}\,,
\eeqn
with two constants $c_f$ and $c_h$.\footnote{The equations of motion can have other, more exotic vacuum solutions,
but generally they are of less physical interest~\cite{Hassan:2014vja}.}
In the following we denote the background metric by $\bar{g}_{\mu\nu}$.
For the above ansatz, the equations of motion reduce to three copies of Einstein's equations,
\begin{subequations}\label{bgeq}
\beqn
\mathcal{G}_{\mu\nu}(\bar{g})+\Big(\Lambda(\beta_n^f,c_f)+\Lambda(\beta_n^h,c_h)\Big)\bar{g}_{\mu\nu}&=&0\,,\\
\mathcal{G}_{\mu\nu}(\bar{g})+\tilde{\Lambda}(\beta_n^f,c_f,\alpha_f)\bar{g}_{\mu\nu}&=&0\,,\\
\mathcal{G}_{\mu\nu}(\bar{g})+\tilde{\Lambda}(\beta_n^h,c_h,\alpha_h)\bar{g}_{\mu\nu}&=&0\,,
\eeqn
\end{subequations}
where $\mathcal{G}_{\mu\nu}(\bar{g})$ is the Einstein tensor of the metric $\bar{g}_{\mu\nu}$ and,
\beqn\label{lambdadef}
\Lambda(\beta_n,c)&=&m^2\left(\beta_0+3c\beta_1+3c^2\beta_2+c^3\beta_3\right)\,,\\
\tilde{\Lambda}(\beta_n,c, \alpha)&=&\frac{m^2}{\alpha^2c^2}
\left(c\beta_1+3c^2\beta_2+3c^3\beta_3+c^4\beta_4\right)\,.
\eeqn
Since the Einstein tensor is scale-invariant one finds the two background conditions,
\beqn\label{eq:cosmconsts}
\Lambda(\beta_n^f,c_f)+\Lambda(\beta_n^h,c_h)=\tilde{\Lambda}(\beta_n^f,c_f,\alpha_f)=\tilde{\Lambda}(\beta_n^h,c_h,\alpha_h)\,.
\eeqn
These determine the proportionality constants $c_f$ and $c_h$ in terms of the parameters of the theory.
For an action with $S_2\times(Z_2)^2$ symmetry, all solutions satisfy,\footnote{Equation (\ref{cratio})
solves the condition $\tilde{\Lambda}(\beta_n^f,c_f,\alpha_f)=\tilde{\Lambda}(\beta_n^h,c_h,\alpha_h)$ 
whenever (\ref{symmcond}) holds. However, if we do not impose (\ref{reflcond}) 
and thus there is only the $S_2$ invariance, there may exist other solutions.}
\beqn\label{cratio}
\alpha_f^2c_f^2=\alpha_h^2c_h^2\,.
\eeqn
Since all cosmological constant contributions in (\ref{bgeq}) are equal, 
we will simply refer to them by the symbol $\Lambda$.

%%%%%%%%%%%%%%%%%%%%%%%%%%%%%%%%%%%%%%%%%%%%%%%%%%%%%%%%%%%
\subsection{Mass eigenstates}\label{sec:masseigenstates}
%%%%%%%%%%%%%%%%%%%%%%%%%%%%%%%%%%%%%%%%%%%%%%%%%%%%%%%%%%%

Next one derives the equations of motion for linear perturbations
of the metrics around the proportional backgrounds, 
$\gmn=\bar{g}_{\mu\nu}+\delta\gmn$,
$\fmn=c_f^2\bar{g}_{\mu\nu}+\delta\fmn$, and 
$\hmn=c_h^2\bar{g}_{\mu\nu}+\delta\hmn$.
The equations will not be diagonal in these fluctuations and hence
the latter do not correspond to the mass eigenstates of the theory.
In order to find the eigenstates one solves the characteristic
equation of the mass matrix. 
For interaction parameters that satisfy the condition (\ref{symmcond}) and (\ref{reflcond}) and thereby
realise the $S_2\times(Z_2)^2$ invariance of the action, 
the canonically normalised eigenstates of the mass matrix assume the following simple 
form,\footnote{
The results of~\cite{Baldacchino:2016jsz} show that
for the mass eigenstates to be of the form (\ref{masseig}) it is sufficient to impose 
${c_h\alpha_h^2}({\beta_1^f + 2c_f\beta_2^f+c_f^3\beta_3^f})=
{c_f\alpha_f^2}({\beta_1^h + 2c_h\beta_2^h+c_h^3\beta_3^h})$, which can be achieved 
by fixing only one parameter.
Our combined conditions (\ref{symmcond}) for the interchange symmetry
and (\ref{reflcond}) for the reflection symmetry (which together
imply (\ref{cratio})) satisfy this constraint but are more restrictive.
Another possibility to obtain (\ref{masseig}) is to require (\ref{symmcond}) alone
and choose a solution satisfying (\ref{cratio}). 
In this case it is the interchange symmetry of 
section~\ref{sec:intsym} that forbids a coupling of $\delta \chi_{\mu\nu}$ to matter
around these backgrounds. 
Imposing the reflection symmetry of section~\ref{sec:reflsym}
in addition ensures that all solutions satisfy~(\ref{cratio}).}
\begin{subequations}\label{masseig}
\beqn
\delta G_{\mu\nu}&=&\frac{\mpl\left( \delta g_{\mu\nu}+\alpha_f^2\delta f_{\mu\nu} +\alpha_h^2\delta h_{\mu\nu}\right)}
{1+\alpha_f^2c_f^2+\alpha_h^2c_h^2}
\,,\\
\delta {M}_{\mu\nu}&=& \frac{{\mpl}\left(\alpha_f^2\delta f_{\mu\nu} +\alpha_h^2\delta h_{\mu\nu}-(\alpha_f^2c_f^2+\alpha_h^2c_h^2)\delta g_{\mu\nu}\right)}
{(1+\alpha_f^2c_f^2+\alpha_h^2c_h^2)\sqrt{\alpha_f^2c_f^2+\alpha_h^2c_h^2}}
\,,\\
\delta \chi_{\mu\nu}&=& \frac{{\mpl}{\alpha_f\alpha_h}\left(\frac{c_f}{c_h}\delta h_{\mu\nu}-\frac{c_h}{c_f}\delta f_{\mu\nu}\right)}
{\sqrt{1+\alpha_f^2c_f^2+\alpha_h^2c_h^2}
\sqrt{\alpha_f^2c_f^2+\alpha_h^2c_h^2}}
\,,
\eeqn
\end{subequations}
where we have defined the physical Planck mass,
\beqn
\mpl\equiv m_g\sqrt{1+\alpha_f^2c_f^2+\alpha_h^2c_h^2}\,.
\eeqn
Note in particular that the state $\delta \chi_{\mu\nu}$ is independent of 
the original fluctuation $\delta \gmn$.
The corresponding mass eigenvalues are given by,
\beqn\label{masses}
\mu_{G}^2=0\,,\qquad
\mu_M^2=\mathcal{A}(1+\alpha_f^2c_f^2+\alpha_h^2c_h^2)m^2\,,\qquad
\mu_\chi^2=\mathcal{A}m^2\,.
\eeqn
with $\mathcal{A}\equiv\frac{2\beta_f^2}{\alpha_f^2}=\frac{2\beta_h^2}{\alpha_h^2}$.
Using the background condition~(\ref{cratio}), the mass eigenstates can be written in the form,
\begin{subequations}\label{masseig2}
\beqn
\delta G_{\mu\nu}&=&\frac{\mpl\left( \delta g_{\mu\nu}+\alpha_f^2\delta f_{\mu\nu} +\alpha_h^2\delta h_{\mu\nu}\right)}
{1+\alpha^2}
\,,\\
\delta {M}_{\mu\nu}&=& \frac{{\mpl}\left(\alpha_f^2\delta f_{\mu\nu} 
+\alpha_h^2\delta h_{\mu\nu}-\alpha^2\delta g_{\mu\nu}\right)}
{\alpha(1+\alpha^2)}
\,,\\
\delta \chi_{\mu\nu}&=& \frac{{\mpl}\left(\alpha_h^2\delta h_{\mu\nu}-\alpha_f^2\delta f_{\mu\nu}\right)}
{\alpha\sqrt{1+\alpha^2}}
\,,
\eeqn
\end{subequations}
where we have defined,
\beqn
\alpha^2\equiv \alpha_f^2c_f^2+\alpha_h^2c_h^2 = 2\alpha_f^2c_f^2= 2\alpha_h^2c_h^2\,.
\eeqn
The quadratic action in terms of mass eigenstates is provided in appendix~\ref{app:cubic}.

Let us comment on some immediate phenomenological implications. 
The spin-2 mass eigenstate $\delta\chi_{\mu\nu}$ does not interact 
directly with the matter sector since the matter coupling contains only 
the fluctuations of the physical metric $\gmn$. 
For instance, the coupling in the quadratic action is of the form,
\beqn
\delta g^{\mu\nu} T_{\mu\nu}
=\frac{1}{\mpl}\left(\delta G^{\mu\nu}-\alpha\,\delta{M}^{\mu\nu}\right)T_{\mu\nu}\,,
\eeqn
where $T_{\mu\nu}$ is the matter stress-energy tensor (here taken to be a small fluctuation
sourcing perturbations on the proportional backgrounds).
Matter couplings involving $\delta\chi_{\mu\nu}$ are forbidden by the global symmetries.
Note furthermore from (\ref{masses}) that the mass of $\delta{M}_{\mu\nu}$ 
is always larger than that of $\delta\chi_{\mu\nu}$ by a factor of $\sqrt{1+\alpha^2}$. 
The particle corresponding to $\delta\chi_{\mu\nu}$
is thus stable against two-body decay into SM particles and other massive spin-2 modes.
In fact, these decay diagrams are also forbidden by the global symmetries, 
as we will explain in section~\ref{sec:cubvert}. Namely, the symmetries forbid all linear couplings 
in $\delta \chi_{\mu\nu}$ and hence higher-order decays 
(such as $\delta\chi\rightarrow 2\delta M \rightarrow 4\,$SM)
with off-shell spin-2 modes are not possible either.
Moreover, in the next section we will demonstrate that
$\delta\chi_{\mu\nu}$ does not decay into massless gravitons and is therefore entirely stable.
A decay of the heavier spin-2 particle into two lighter ones, 
$\delta{M}\rightarrow\delta\chi\delta\chi$,
is possible provided that $\mu_{ M}>2\mu_{ \chi}$. Since
 $\mu_{ M}^2=(1+\alpha^2)\mu^2_{ \chi}$
this requires $\alpha^2>3$. 
A decay into one massive and any number of massless particles, 
$\delta{M}\rightarrow\delta\chi\delta G\delta G\hdots$,
is again forbidden by the global symmetries.

In section~\ref{sec:gen} we will see that all of these results naturally generalise
to theories with $N$ satellite fields and symmetry $S_N\times(Z_2)^N$.

%%%%%%%%%%%%%%%%%%%%%%%%%%%%%%%%%%%%%%%%%%%%%%%%%%%%%%%%%%%%%%%%%%%%%%
\section{Perturbative expansion of the action}\label{sec:perturbact}
%%%%%%%%%%%%%%%%%%%%%%%%%%%%%%%%%%%%%%%%%%%%%%%%%%%%%%%%%%%%%%%%%%%%%%

In this section we discuss the perturbative expansion of the trimetric action
with maximal symmetry in terms of mass eigenstates.
Inverting the relations (\ref{masseig2}), we can express the original metric fluctuations in 
terms of the mass eigenstates as follows,
\begin{subequations}\label{invme}
\beqn
\delta g_{\mu\nu} &=&\frac{1}{\mpl}\left(\delta G_{\mu\nu}-\alpha\,\delta{M}_{\mu\nu}\right)\,,\\
\delta f_{\mu\nu} &=&\frac{\alpha}{2\alpha_f^2\mpl}\left(\alpha\delta G_{\mu\nu}-\sqrt{1+\alpha^2}
\delta \chi_{\mu\nu}
+\delta M_{\mu\nu}\right)\,,\\
\delta h_{\mu\nu} 
&=&\frac{\alpha}{2\alpha_h^2\mpl}\left(\alpha\delta G_{\mu\nu}+
\sqrt{1+\alpha^2}
\delta \chi_{\mu\nu}
+\delta M_{\mu\nu}\right)\,.
\eeqn
\end{subequations}
These expression can now be used to compute the interaction vertices
of mass eigenstates to all orders in the trimetric action~(\ref{triact})
with parameters satisfying the conditions (\ref{symmcond}) and (\ref{reflcond}).

For $\alpha< 1$, the scale suppressing higher-order vertices of $\delta M$ and $\delta\chi$ is $\alpha\mpl$.
The perturbative structure is thus formally the same as in bimetric theory where $\alpha$ was simply the ratio of two
Planck masses. 
It was discussed in detail in \cite{Babichev:2016bxi} that in this case the perturbative expansion is valid for energies 
smaller than $\alpha\mpl$.

%%%%%%%%%%%%%%%%%%%%%%%%%%%%%%%%%%%%%%%%%%%%%%%%%%%%%%%%%%%%%%%%%%%%%%%%%%%%%%%%%
\subsection{Nonlinear massless field}\label{sec:masslf}
%%%%%%%%%%%%%%%%%%%%%%%%%%%%%%%%%%%%%%%%%%%%%%%%%%%%%%%%%%%%%%%%%%%%%%%%%%%%%%%%%

Just like in the bimetric case (see~\cite{Babichev:2016bxi}), 
it is possible to define a nonlinear massless metric,
\beqn
G_{\mu\nu}=\bar{g}_{\mu\nu}+\frac{1}{\mpl}\delta G_{\mu\nu}\,.
\eeqn
In terms of this, we can then write the full metrics $\gmn$, $\fmn$ and $\hmn$ as follows,
\begin{subequations}\label{mesnl}
\beqn
g_{\mu\nu}&=&G_{\mu\nu}-\frac{\alpha}{\mpl}\delta{M}_{\mu\nu}\,,\\
f_{\mu\nu}&=&\frac{\alpha^2}{2\alpha_f^2}G_{\mu\nu}+\frac{\alpha}{2\alpha_f^2\mpl}\left(\delta M_{\mu\nu}
-\sqrt{1+\alpha^2}\,\delta \chi_{\mu\nu}\right)
\,,\\
h_{\mu\nu}&=&\frac{\alpha^2}{2\alpha_h^2}G_{\mu\nu}
+\frac{\alpha}{2\alpha_h^2\mpl}\left(\delta M_{\mu\nu}
+\sqrt{1+\alpha^2}\,\delta \chi_{\mu\nu}\right)\,.
\eeqn
\end{subequations}
The field $G_{\mu\nu}$ is massless in the following sense.
If we insert the expressions (\ref{mesnl}) back into (\ref{triact}) and 
formally neglect the massive modes by setting $\delta M_{\mu\nu}=\delta\chi_{\mu\nu}=0$, 
we recover the Einstein-Hilbert action for $G_{\mu\nu}$, 
\beqn
\left.S[g,f,h]\right|_{\delta M=\delta\chi=0}=\mpl^2\int\dd^4x~\sqrt{|G|}\,\left(R(G)-2\Lambda\right)\,,
\eeqn
where $\Lambda$ is identical to the cosmological constant of the background metric $\bar{g}_{\mu\nu}$. 
The self-interactions of $G_{\mu\nu}$ are exactly the same as those of the massless tensor
field in GR. Even at the nonlinear level, we can therefore interpret the fluctuation 
$\delta G_{\mu\nu}$ as the massless field mediating the long-ranged gravitational force,
and the metric $G_{\mu\nu}$ sets the geometry in which the massive spin-2 modes propagate.

%%%%%%%%%%%%%%%%%%%%%%%%%%%%%%%%%%%%%%%%%%%%%%%%%%%%%%%%%%%%%%%%%%%%%%%%%%%%%%%%%
\subsection{Vertices linear in both massive modes}\label{sec:linv}
%%%%%%%%%%%%%%%%%%%%%%%%%%%%%%%%%%%%%%%%%%%%%%%%%%%%%%%%%%%%%%%%%%%%%%%%%%%%%%%%%

Next we study the vertices linear in the massive modes, i.e.~we expand the trimetric action~(\ref{triact})
to linear order in $\delta M_{\mu\nu}$ and $\delta \chi_{\mu\nu}$, formally keeping all orders of $G_{\mu\nu}$.
This gives an all-order result but neglects vertices including more than one massive mode. 
The contributions from the Einstein-Hilbert terms are,
\begin{align}
&m_g^2\int d^4x\left.\tfrac{\delta\sqrt{|g|}\,R(g)}{\delta g_{\mu\nu}}\right|_{g=G}
\left(\delta\gmn+\alpha_f^2\delta\fmn+\alpha_h^2\delta\hmn\right)\\
=~
%~&\mpl^2\int\dd^4x~\sqrt{G}\,R(G)\\
&\tfrac{m_g^2}{\mpl}\int d^4x\left.\tfrac{\delta\sqrt{|g|}\,R(g)}{\delta g_{\mu\nu}}\right|_{g=G}
\left[\Big(\tfrac{\alpha_f^2c_f^2}{\alpha}+\tfrac{\alpha_h^2c_h^2}{\alpha}-\alpha\Big)\delta M_{\mu\nu}
+\tfrac{\sqrt{1+\alpha^2}}{\alpha}\Big(\alpha_h^2c_h^2-\alpha_f^2c_f^2\Big)\delta \chi_{\mu\nu}\right]\,.
\end{align}
Since $\alpha^2=\alpha_h^2c_h^2+\alpha_f^2c_f^2$ and $\alpha_h^2c_h^2=\alpha_f^2c_f^2$,
the round brackets in the last line both vanish and hence there are no vertices linear in the massive modes
coming from the Einstein-Hilbert terms.
Taylor expansion of the interaction potential $\sqrt{|g|}\,V=\sqrt{|g|}\,(V(g,f;\beta_n^f)+V(g,h;\beta_n^h))$ to first order around the proportional backgrounds 
gives rise to the following couplings,
\begin{align}
%\nn ~&m^2\left[\sqrt{g}\Big(V(g,f;\beta^f_n)+V(g,h;\beta^h_n)\Big)\right]_{f=h=g=G}\\
&~m^2\left.\tfrac{\pa(\sqrt{|g|}\,V)}{\pa\delta \chi_{\rho\sigma}}\right|_{f=h=g=G}\delta \chi_{\rho\sigma}
+m^2\left.\tfrac{\pa(\sqrt{|g|}\,V)}{\pa\delta M_{\rho\sigma}}\right|_{f=h=g=G}\delta M_{\rho\sigma}
\nn\\
 =
%~&(1+\alpha_f^2+\alpha_h^2)\tilde{\Lambda}(\beta^f_n,c,\alpha_f)\sqrt{G}\\
\nn &\left[\tfrac{\pa(\sqrt{|g|}V(g,f;\beta^f_n))}{\pa f^{\mu\nu}}\tfrac{\pa f^{\mu\nu}}{\pa\delta \chi_{\rho\sigma}}
+\tfrac{\pa(\sqrt{|g|}V(g,h;\beta^h_n))}{\pa h^{\mu\nu}}
\tfrac{\pa h^{\mu\nu}}{\pa\delta \chi_{\rho\sigma}}\right]_{f=h=g=G}\delta \chi_{\rho\sigma}\nn\\
\nn &+\left[\tfrac{\pa(\sqrt{|g|}V)}{\pa g^{\mu\nu}}\tfrac{\pa g^{\mu\nu}}{\pa\delta{M}_{\rho\sigma}}
+\tfrac{\pa(\sqrt{|g|}V(g,f;\beta^f_n))}{\pa f^{\mu\nu}}\tfrac{\pa f^{\mu\nu}}{\pa\delta{M}_{\rho\sigma}}
+\tfrac{\pa(\sqrt{|g|}V(g,h;\beta^h_n))}{\pa h^{\mu\nu}}\tfrac{\pa h^{\mu\nu}}{\pa\delta{M}_{\rho\sigma}}
\right]_{f=h=g=G}\delta M_{\rho\sigma}\\
\nn \equiv~&
%(1+\alpha_f^2+\alpha_h^2)\tilde{\Lambda}(\beta^f_n,c,\alpha_f)\sqrt{G}
\tfrac{\sqrt{1+\alpha^2}}{2\alpha}\tilde{\Lambda}(\beta^f_n,c_f,\alpha_f)
\Big(\alpha_h^2c_h^2-\alpha_f^2c_f^2\Big) \sqrt{|G|}\,G^{\rho\sigma}\delta \chi_{\rho\sigma}\nn\\
&+\tfrac{1}{2}\tilde{\Lambda}(\beta^f_n,c_h,\alpha_f)
\Big(\tfrac{\alpha_f^2c_f^2}{\alpha}+\tfrac{\alpha_h^2c_h^2}{\alpha}-\alpha\Big)
\sqrt{|G|}\, G^{\rho\sigma}\delta M_{\rho\sigma}\,.
\end{align}
where we have evaluated the variations of the potential terms on the background 
and used (\ref{eq:cosmconsts}) and (\ref{mesnl}).
The expressions in the round brackets again vanish and 
the interaction potential does not contain linear fluctuations in the massive modes either.

%%%%%%%%%%%%%%%%%%%%%%%%%%%%%%%%%%%%%%%%%%%%%%%%%%%%%%%%%%%%%%%%%%%%%%%%%%%%%%%%
\subsection{Cubic vertices}\label{sec:cubvert}
%%%%%%%%%%%%%%%%%%%%%%%%%%%%%%%%%%%%%%%%%%%%%%%%%%%%%%%%%%%%%%%%%%%%%%%%%%%%%%%%%

We now obtain the expression for the perturbed trimetric action to cubic order in
mass eigenstates.
Our full and explicit results can be found in appendix~\ref{app:cubic}.
In Table~\ref{tbl:vertices} we collect the pre-factors of cubic interaction vertices, 
paying particular attention to the dependence on $\alpha$, but neglecting numerical factors. 
Dimensionless factors always multiply kinetic interactions.
%\begin{table}[htdp]
\begin{table}[htp]
\caption{Prefactors of cubic vertices; all suppressed by $\mpl^{-1}$.}
\begin{center}
\begin{tabular}{|c|c|c|c|c|c|}
\hline
%&&&&&&&&\\
&& $\delta\chi^3$, $\dG^2\dM$, &&&\\
$\dG^3$ &$\dM^3$ &$\dG^2\delta\chi$, $\dM^2\delta\chi$ 
& $\dM^2\dG$ &$\delta\chi^2\dG$ & $\delta\chi^2\dM$\\
\hline
&&&&&\\
$1, \Lambda$ &$\frac{1-\alpha^2}{\alpha}\cdot (1, \Lambda, \mu_M^2)$
&$0$
& $1,\Lambda, \mu_M^2$ &$1,\Lambda, \mu_\chi^2$ & $\frac{1}{\alpha}\cdot (1, \Lambda,\mu_M^2, \mu_\chi^2)$\\
&&&&&\\
\hline 
\end{tabular}
\end{center}
\label{tbl:vertices}
\end{table}%

We notice a couple of distinct features.
\begin{itemize}
\item The cubic self-interactions of the massless mode $\dG_{\mu\nu}$ are exactly those of GR, 
which is consistent with the all-order results of section~\ref{sec:masslf}.
\item As expected from the general results derived in section~\ref{sec:linv}, 
there are no terms linear in both $\delta M_{\mu\nu}$ and $\delta\chi_{\mu\nu}$.
This explicitly confirms that the cubic vertices do not give rise to diagrams 
describing decay into massless gravitons.
\item 
The cubic action does not contain terms with odd powers of $\delta\chi_{\mu\nu}$.
In fact, this holds to all orders in perturbation series which can be seen as follows.
Under the interchange symmetry $\alpha_f^2\fmn\leftrightarrow\alpha_h^2\hmn$, which leaves
the action invariant, this mode transforms as $\delta\chi_{\mu\nu}\rightarrow -\delta\chi_{\mu\nu}$,
whereas the other two mass eigenstates are invariant,
$\delta M_{\mu\nu}\rightarrow \delta M_{\mu\nu}$ and 
$\delta G_{\mu\nu}\rightarrow \delta G_{\mu\nu}$.
Hence, a term with odd powers of $\delta\chi_{\mu\nu}$ would transform into its negative and 
spoil the invariance of the action. It can therefore not exist.
Consequently a decay into one massive and any number of massless particles, 
$\delta{M}\rightarrow\delta\chi\delta G\delta G\hdots$,
is forbidden because it requires a vertex that is linear in $\delta\chi_{\mu\nu}$. 
\item The vertices with two massive and one massless mode do not depend on $\alpha$. 
This is expected because they will enter the expression for the gravitational stress-energy
tensor, which coincides with the Noether stress-energy defined for the
($\alpha$-independent) quadratic action in flat space~\cite{Leclerc:2005na}.
\item All of the cubic self-interactions of $\delta M_{\mu\nu}$ are enhanced for both 
small and large $\alpha$, but with opposite sign. 
Interestingly, they all vanish for $\alpha^2=1$.
\item All $\delta M\delta\chi\delta\chi$ terms are enhanced for small $\alpha$.  
On the other hand, while everything is naturally Planck scale suppressed, with a large value for
$\alpha$ the suppression of these terms would be even stronger. 
\end{itemize}

%%%%%%%%%%%%%%%%%%%%%%%%%%%%%%%%%%%%%%%%%%%%%%%%%%%%%%%%%%%
\subsection{Generalisation to multiple fields}\label{sec:gen}
%%%%%%%%%%%%%%%%%%%%%%%%%%%%%%%%%%%%%%%%%%%%%%%%%%%%%%%%%%%

The mass spectrum of a general multimetric theory is rather complicated.
For a theory of the form~(\ref{multi}) it has been worked out explicitly in~\cite{Baldacchino:2016jsz}.
The mass eigenstates are linear combinations of the metric fluctuations $\delta\gmn$ and
$\delta\fmn^{(i)}$ around a maximally symmetric background solution.
The spectrum always contains one massless mode $\delta G_{\mu\nu}$ and $N$
massive modes $\delta \chi^{(i)}_{\mu\nu}$ with masses~$\mu_i$.

From the results of~\cite{Baldacchino:2016jsz} we can derive a feature shared
by all models with at least one discrete interchange symmetry of section~\ref{sec:intsym}
and two reflection symmetries of section~\ref{sec:reflsym}:
When the conditions (\ref{symmcond}) and (\ref{reflcond}) are imposed on the same two parameter sets,
then in one of the massive states $\delta \chi^{(i)}_{\mu\nu}$ 
the coefficient in front of $\delta\gmn$ vanishes and hence this massive mode
drops out of the matter coupling.

In particular, for the action (\ref{maxsym}) with maximal symmetry $S_N\times(Z_2)^N$, 
the proportional backgrounds $\fmn^{i}=c_i^2\gmn$ satisfy $c_i^2/c_j^2=\alpha_j^2/\alpha_i^2$
and the mass spectrum takes on the form,
\begin{subequations}\label{multeig}
\beqn
\delta G_{\mu\nu}&=&\frac{\mpl}{1+\alpha^2}\left( \delta g_{\mu\nu}
+\sum_{k=1}^N\alpha_k^2\delta f^{(k)}_{\mu\nu} \right)\,,\\
\delta {M}_{\mu\nu}&=& \frac{{\mpl}}{(1+\alpha^2)\alpha}\left(\sum_{k=1}^N\alpha_k^2\delta f^{(k)}_{\mu\nu}
-\alpha^2\delta g_{\mu\nu}\right)\,,\\
\delta \chi^{(i)}_{\mu\nu}&=& \frac{\mpl}{\alpha\sqrt{1+\alpha^2}}
\left(\alpha_{i}^2\delta f^{(i)}_{\mu\nu}-\alpha_{i+1}^2\delta f^{(i+1)}_{\mu\nu}\right)\qquad i=1,\dots, N-1\,.
\eeqn
\end{subequations}
with $\alpha^2=\sum_{k=1}^N\alpha_k^2c_k^2$ and $\mpl=\sqrt{1+\alpha^2}\, m_g$. 
One mode (which we have labelled $\delta M_{\mu\nu}$) retains its dependence on $\delta\gmn$ because
imposing the full $S_N$ invariance gives only $(N-1)$ conditions of the form (\ref{symmcond}). 
Since the fields $\delta \chi^{(i)}_{\mu\nu}$ do not depend on the fluctuation $\delta\gmn$ of
the physical metric, they do not show up in the coupling to matter. 
Their masses $\mu_i$ are all
equal to each other and related to the mass ${\mu}_M$ of $\delta M_{\mu\nu}$ through,
\beqn
\mu_i=\mu_j=\frac{{\mu}_M}{\sqrt{1+\alpha^2}}\qquad \forall~i,j=1,\dots, N-1\,.
\eeqn
As in the trimetric case, we derive from (\ref{multeig}) that,
\beqn\label{gitome}
\delta g_{\mu\nu} =\frac{1}{\mpl}\left(\delta G_{\mu\nu}-\alpha\,\delta{M}_{\mu\nu}\right)\,.
\eeqn
Matter thus only interacts with the massless and with the heaviest spin-2 field
while the $(N-1)$ massive spin-2 particles corresponding to the modes $\delta \chi^{(i)}_{\mu\nu}$ 
cannot decay into SM particles. Moreover, they all have equal mass and are lighter than 
$\delta {M}_{\mu\nu}$, which forbids the two-body decay into massive spin-2 particles. 
Higher order decay channels are again forbidden by the global symmetries which ensure the
absence of terms linear in $\delta \chi^{(i)}_{\mu\nu}$.
The only remaining decay channel would be into massless gravitons
but the generalisation of the calculation in section~\ref{sec:linv} shows that there 
are no couplings giving rise to such decay diagrams.

%%%%%%%%%%%%%%%%%%%%%%%%%%%%%%%%%%%%%%%%%%%%%%%%%%%%%%%%%%%
\section{Dark matter phenomenology}\label{sec:dm}
%%%%%%%%%%%%%%%%%%%%%%%%%%%%%%%%%%%%%%%%%%%%%%%%%%%%%%%%%%%

The absolute stability of the lightest mode $\delta \chi_{\mu\nu}$ motivates us
to consider it as a possible candidate for the DM particle. 
Before exploring the phenomenological implications of this idea,
we make a few comments on how we restrict the ranges for the parameters of 
maximally symmetric trimetric theory.

%%%%%%%%%%%%%%%%%%%%%%%%%%%%%%%%%%%%%%%%%%%%%%%%%%%
\subsection{Parameter regions of interest}
%%%%%%%%%%%%%%%%%%%%%%%%%%%%%%%%%%%%%%%%%%%%%%%%%%%

The particle corresponding to the massive spin-2 mode $\delta \chi_{\mu\nu}$ is completely stable since
it does not decay into massive spin-2 fields $\delta M_{\mu\nu}$,
nor massless gravitons $\delta G_{\mu\nu}$ and it does not couple to Standard Model fields. 
Even without any global symmetries in the action, 
the lightest spin-2 mode would not decay into $\delta G_{\mu\nu}$, since this coupling 
does not arise in any diffeomorphism invariant theory.
The decay into SM matter on the other hand is forbidden by the global 
$S_2\times(Z_2)^2$ symmetry of the theory. We expect (but do not prove) that both the diffeomorphism
invariance and the global symmetry are stable under quantum corrections.\footnote{Note that there is no
obvious symmetry protecting the vanishing SM matter couplings of the satellite metrics $\fmn$ and $\hmn$. 
Such couplings would reintroduce ghosts at the quantum level. This is an unresolved problem in bi-
and multimetric theory which we will not address here.} 
Provided that this is correct,
no parameters need to be tuned in order to guarantee the stability of the particle corresponding
to $\delta \chi_{\mu\nu}$.

The Planck mass $\mpl$ is 
known to be large, on the order of $10^{18}$\,GeV, giving rise to feeble gravitational interactions.
This creates the well-known hierarchy between the Planck scale and the weak scale.  
Moreover, the cosmological constant $\Lambda$ is measured to be very small, $\Lambda\simeq 10^{-122}\mpl^2$.
We do not attempt to address these hierarchy problems here, but tune the values of $\mpl$ and $\Lambda$ 
to match the gravitational and cosmological observations.

We need to make an assumption on the dimensionless parameter
$\alpha$ that parameterises the interaction strengths of massive spin-2 modes.
It is obvious that the limit $\alpha^2=\alpha_f^2c_f^2 +\alpha_h^2c_h^2\rightarrow 0$ 
is the generalisation of the GR limit $\alpha\rightarrow 0$ in bimetric 
theory~\cite{Baccetti:2012bk, Hassan:2014vja}. 
More precisely, in the bimetric case it is known that the parameter $\alpha$
and the spin-2 mass $m$ together control the deviations from GR~\cite{Babichev:2016bxi}. Demanding
$\alpha\ll 1$ ensures that these deviations are small for a large range of spin-2 masses (and
in particular small ones). It is easy to see that the situation will be similar in the
presence of multiple massive spin-2 fields and we take this as a motivation to begin
our investigations by focusing on values $\alpha^2=\alpha_f^2c_f^2+\alpha_h^2c_h^2< 1$ in this work. 
We will comment on the interesting implications of larger values for $\alpha$ 
in section~\ref{sec:discussion}.

One important implication of the assumption $\alpha< 1$ is that 
the masses of the spin-2 modes in (\ref{masses}) are of the same order, 
$\mu_M\simeq \mu_\chi$. 
In this case $\delta M_{\mu\nu}$ cannot decay into $\delta \chi_{\mu\nu}$ since $2\mu_\chi>\mu_M$.
We take $\beta^f_2=\alpha_h^2\beta_2^h/\alpha_f^2$ to be of order unity\footnote{This assumption
is without loss of generality because the scale of the $\beta_2$ can always be absorbed into $m$.}
 and hence $\mu_M\simeq\mu_\chi\simeq m$. 
The heaviest mode can of course still decay into 
Standard Model fields and its matter coupling is of the same form as in bimetric theory. 
It then follows from the results of \cite{Babichev:2016bxi} that if $\delta M_{\mu\nu}$ makes up 
(part of) the observed dark matter density, its non-observation in indirect detection experiments
requires $10^{-15}\lesssim\alpha\lesssim 10^{-12}$ and $m\simeq 1-100$\,TeV.\footnote{The 
enhanced self-interactions of the DM particle (proportional to $1/\alpha$ at cubic order) can result 
in a thermalisation of the spin-2 sector, and the DM abundance could be produced 
via a ``dark freeze-out"~\cite{Chu:2017msm}.
For this to be effective, $\alpha$ needs to be as small as $10^{-20}$ and the mass can be lowered to 1\,MeV.
We do not take self-interactions into account in what follows
since their impact is expected to be weaker due to the larger values for $\alpha$ in our new scenarios.} 
In other words, the bimetric DM scenario with freeze-in production mechanism
can also be realised in the presence of more than one massive spin-2 particle and, since the 
lighter field $\delta\chi_{\mu\nu}$ is not observable in indirect detection experiments, 
the phenomenology will essentially be the same.

Interestingly, the multimetric theory leaves us with more options.
Namely we can require that $\delta M_{\mu\nu}$
does not contribute to the observed dark matter density because it has decayed since the end of inflation.
This basically provides us with the reversed stability bound derived in~\cite{Babichev:2016bxi},
\beqn\label{revstab}
\alpha^{2/3}\mu_M > 0.13\,\mathrm{GeV}\,.
\eeqn
For fixed spin-2 mass, this translates into a lower bound on $\alpha$.
For instance, for $\mu_M\simeq 1$\,TeV we need $\alpha> 10^{-6}$.
Without an abundance of $\delta M_{\mu\nu}$ particles, there are no constraints from
indirect detection experiments forcing $\alpha$ to be small.
In fact, as we will mention in section~\ref{sec:sa}, such constraints  
seem to favour larger values for $\alpha$.
Let us emphasise once more that this has no effect on the stability of the lighter spin-2 field
$\delta \chi_{\mu\nu}$ which is our dark matter candidate.
In what follows we will always impose the bound~(\ref{revstab}), keeping in mind
that there exists another region in parameters space with much smaller $\alpha$
where the analogue of the bimetric DM scenario could be at work.

%%%%%%%%%%%%%%%%%%%%%%%%%%%%%%%%%%%%%%%%%%%%%%%%%%%%%%%%
\subsection{Production mechanisms for $\alpha<1$}
%%%%%%%%%%%%%%%%%%%%%%%%%%%%%%%%%%%%%%%%%%%%%%%%%%%%%%%%

The assumptions motivated in the previous section can be summarised as,
%\begin{subequations}
\begin{align}
\left(\frac{0.13\,\mathrm{GeV}}{\mu_M}\right)^{3/2}<\alpha<1\,,\qquad 
\beta^f_2=\frac{\alpha_h^2\beta_2^h}{\alpha_f^2}\simeq 1\,,\qquad
\Rightarrow~~\mu_M\simeq \mu_\chi\simeq m\,.
\end{align}
%\end{subequations}
DM related observations now essentially constrain the remaining
parameters $\alpha$ and $m$.

In~\cite{Babichev:2016bxi} it was explained why the freeze-out mechanism cannot
be responsible for the production of massive spin-2 dark matter. This is due to
the Planck suppression of the coupling between $\delta M_{\mu\nu}$ and matter fields,
which does not allow the two sectors to reach thermal equilibrium in the early Universe.  
The same issue arises in multimetric theory (at least for $\alpha<1$)
and we have to rely on different production mechanisms for spin-2 dark matter. In the following
we argue that both freeze-in~\cite{Hall:2009bx} and 
gravitational particle production~\cite{Chung:1998zb, Kuzmin:1998uv, Chung:2004nh} 
can yield the observed dark matter abundance.

%%%%%%%%%%%%%%%%%%%%%%%%%%%%%%%%%%%%%%%%%%%%%%%%%%%%%%
\subsubsection{Scenario I: Freeze-in production}
%%%%%%%%%%%%%%%%%%%%%%%%%%%%%%%%%%%%%%%%%%%%%%%%%%%%%%

Even though the massive spin-2 sector does not attain thermal equilibrium with
matter particles in the early Universe, a relic density can be produced through a slow 
“leakage” from the thermal bath, resulting in a non-thermalised DM sector~\cite{Hall:2009bx}.
In our case two Standard Model particles from the thermal bath annihilate and 
produce a pair of massive spin-2 particles via $s$-channel exchange of 
$\dG_{\mu\nu}$ or $\dM_{\mu\nu}$.  
This very slow process is never counterbalanced by the opposite one because the 
abundance of massive spin-2 always remains well below the thermal one.
The production can take place during reheating or in the radiation 
dominated era thereafter~\cite{Garny:2015sjg, Tang:2016vch}.

The end products of the $s$-channel production are pairs of spin-2 particles corresponding
to either $\dM_{\mu\nu}$ or $\delta\chi_{\mu\nu}$ (with approximately equal masses $m$).
A $\dM\delta\chi$ pair is not possible due to the absence of couplings linear in 
$\delta\chi_{\mu\nu}$ (c.f.~the discussion in section~\ref{sec:cubvert}).
The dominant Feynman diagrams contain the linear matter coupling and a cubic
interaction vertex (here we use the approximation $1+\alpha^2\simeq 1$):
\begin{center}
{\renewcommand{\arraystretch}{0.1}
\begin{tabular}{p{0.45\textwidth} p{0.45\textwidth}}
\begin{center}
\begin{tikzpicture}[thick,
	level/.style={level distance=1.8cm},
	level 2/.style={sibling distance=2.6cm}]
	\coordinate
		child[grow=left]{
			child {
				node {SM}
				edge from parent [sm]
 			}
 			child {
				node {SM}
				edge from parent [sm]
			}
			edge from parent [graviton] node [above=7pt] {$\dG$}
            node [left=50pt] {$\frac{1}{\mpl}$}
            node [right=50pt] {$\frac{1}{\mpl}$}
		}
		child[grow=right, level distance=0pt] {
			child {
				node {$\dM$,\,$\delta\chi$}
				edge from parent [Mgraviton]
			}
			child {
				node {$\dM$,\,$\delta\chi$}
				edge from parent [Mgraviton]
			}
		};
\end{tikzpicture}
\end{center}
&
\begin{center}
\begin{tikzpicture}[thick,
	level/.style={level distance=1.8cm},
	level 2/.style={sibling distance=2.6cm}]
	\coordinate
		child[grow=left]{
			child {
				node {SM}
				edge from parent [sm]
 			}
 			child {
				node {SM}
				edge from parent [sm]
			}
			edge from parent [Mgraviton] node [above=7pt] {$\dM$}
            node [left=50pt] {$\frac{\alpha}{\mpl}$}
            node [right=50pt] {$\frac{1}{\alpha\mpl}$}
		}
		child[grow=right, level distance=0pt] {
			child {
				node {$\dM$,\,$\delta\chi$}
				edge from parent [Mgraviton]
			}
			child {
				node {$\dM$,\,$\delta\chi$}
				edge from parent [Mgraviton]
			}
		};
\end{tikzpicture}
\end{center}
\\
\begin{center}
{\small Production of a $\dM\dM$ or a $\delta\chi\delta\chi$ pair via $s$-channel exchange of $\dG$. }
\end{center}
&
\begin{center}
{\small Production of a $\dM\dM$ or a $\delta\chi\delta\chi$ pair via $s$-channel exchange of $\dM$. }
\end{center}
\end{tabular}}
\end{center}
They all contribute equally and are independent of $\alpha$.
In principle now both the $\dM_{\mu\nu}$ and $\delta\chi_{\mu\nu}$ particles could contribute to the 
DM abundance. However, as explained in the previous subsection, if the $\dM_{\mu\nu}$ particle
made up part of DM then this would force $\alpha$ to be very tiny and essentially bring us back to the 
bimetric scenario. We therefore restrict to parameters that satisfy the bound (\ref{revstab}) such
that most of the $\dM_{\mu\nu}$ particles have decayed over the age of the universe and we are
left with the entirely stable DM particle $\delta\chi_{\mu\nu}$ alone.

\begin{figure}[h]
    \begin{center}
    \includegraphics[width=370pt]{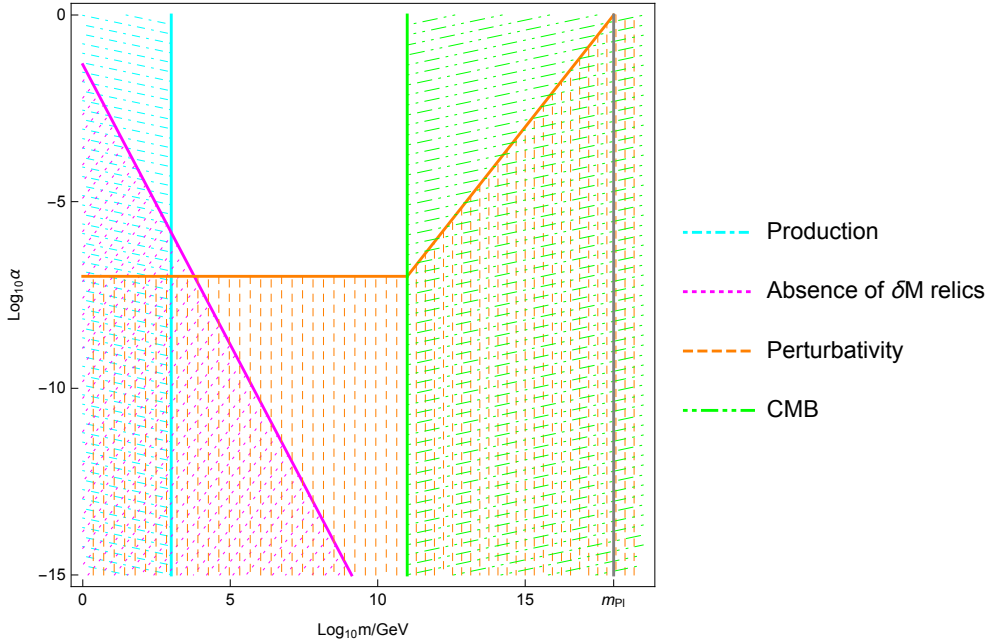}
    \caption{Excluded regions in the $m$-$\alpha$-plane for freeze-in production.}
    \end{center}
    \end{figure}

The allowed mass region for $m_\chi\simeq m$ is the same as in the bimetric case
before imposing the constraints from indirect detection~\cite{Babichev:2016bxi},
\beqn
1\,\mathrm{TeV}\lesssim m \lesssim 10^{11}\,\mathrm{GeV}\,.
\eeqn
The lower bound stems from demanding the observed DM abundance to be generated during 
either reheating or radiation domination. It is sensitive to the efficiency of the reheating
process, $\epsilon_\mathrm{rh}=\frac{\pi^2g_*T^4_{\mathrm{rh}}}{90\mpl^2H_e^2}\leq 1$, with 
reheating temperature $T_{\mathrm{rh}}$, Hubble scale at the end of inflation $H_e$ and $g_*=106.75$
being the number of relativistic degrees of freedom during reheating.
The displayed bound corresponds to maximal efficiency. For lower values of $\epsilon_\mathrm{rh}$,
the bound on the mass moves to higher values. The upper bound is essentially
a constraint on the scale of inflation from non-observation of tensor modes in the CMB.

The allowed values for $\alpha$ are obtained by demanding the absence of a relic density
of $\delta M_{\mu\nu}$ particles and validity of the perturbative expansion. They can
be read off from Figure~3. The perturbativity bound is a combination of demanding
$m<\alpha \mpl$ and $T_\mathrm{rh}<\alpha \mpl$, where for the latter we have assumed
the minimal value required for efficient production, 
$T_\mathrm{rh}\simeq 10^{-7}\mpl$~\cite{Tang:2016vch}.

%%%%%%%%%%%%%%%%%%%%%%%%%%%%%%%%%%%%%%%%%%%%%%%%%%%%%%%%%%%%%%%%
\subsubsection{Scenario II: Gravitational production}
%%%%%%%%%%%%%%%%%%%%%%%%%%%%%%%%%%%%%%%%%%%%%%%%%%%%%%%%%%%%%%%%
Another possible origin of a relic density of stable, massive spin-2 particles
is the non-adiabatic transition of the Universe out of its de-Sitter phase at the end of inflation.  
This change in the cosmological expansion induces a non-adiabatic change in
the frequencies of the Fourier modes defining the particles. This in turn leads to mixing between
modes with positive and negative frequency and thus to quantum creation of particles.
The mechanism, known as gravitational particle production, is effective only for very heavy masses, 
for details see~\cite{Chung:1998zb, Kuzmin:1998uv, Chung:2004nh}.
\begin{figure}[h]
    \begin{center}
    \includegraphics[width=370pt]{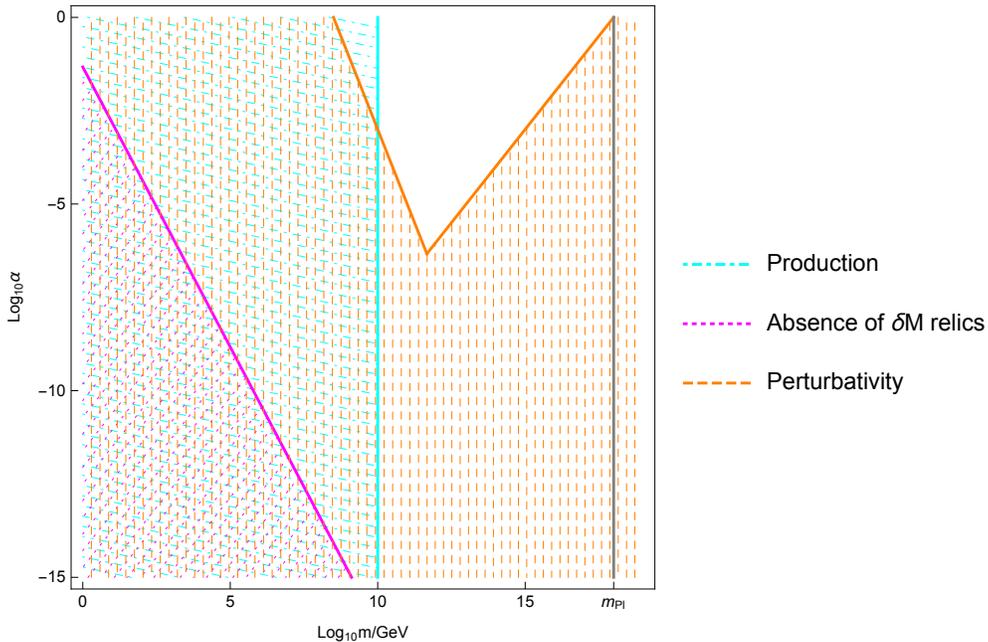}
    \caption{Excluded regions in the $m$-$\alpha$-plane for gravitational particle production.}
    \end{center}
    \end{figure}
The bimetric scenario of~\cite{Babichev:2016bxi} excluded it as a
possible mechanism generating the non-thermal relic density by invoking constraints on 
isocurvature perturbations. Namely, these translated into a lower limit for the spin-2 mass,
\beqn\label{massbgp}
m\gtrsim 10^{10}\,\mathrm{GeV}\,,
\eeqn
which together with constraints from indirect detection experiments resulted in a very low upper 
bound on the analogue of the parameter $\alpha$ in the bimetric case. 
Such a small $\alpha$ was inconsistent with perturbativity. 
In contrast, in the multimetric case, the DM particle does not decay and hence the constraints from indirect detection
are irrelevant, provided that (\ref{revstab}) holds. Together, (\ref{revstab}) and (\ref{massbgp}) now give
a very weak lower bound on $\alpha$ instead,
namely $\alpha\gtrsim 10^{-16}$. 
The requirement of perturbativity, $m\simeq\mu_M<\alpha\mpl$ is of more relevance in this case, because it 
requires $\alpha\gtrsim 10^{-8}$ for the lowest possible spin-2 mass.
The constraints on the scenario are summarised in Figure~4.

Note that the bound on the mass coming from isocurvature perturbations is in general sensitive to the 
Hubble scale at the end of inflation $H_e$. More precisely, it can be estimated as 
$m/H_e\gtrsim 5$ which, after using the value of the observed DM abundance, 
results in~\cite{Babichev:2016bxi},
\beqn
m\gtrsim 10^{14}\left(\frac{10^7\mathrm{GeV}}{T_{\mathrm{rh}}}\right)^{1/2}\mathrm{GeV}\,.
\eeqn
The bound given in (\ref{massbgp}) then assumes a maximal value for
the reheating temperature $T_{\mathrm{rh}}$ given by 
$T_{\mathrm{rh}}\simeq(H_e\mpl)^{1/2}\simeq 10^{16}r_s^{1/4}$GeV 
with tensor-to-scalar ratio $r_s\lesssim 0.7$.
The perturbativity bound in Figure~4 again corresponds to demanding both $m$ and $T_{\mathrm{rh}}$
to be smaller than $\alpha\mpl$.

\vspace{10pt}
In principle, both production mechanisms are of course expected to be at work simultaneously. 
However, in most of the parameter space, one mechanism dominates strongly over the other, 
which is why we treated them separately. Only in the region around 
masses of $10^{10}-10^{11}\,\mathrm{GeV}$, both mechanism could in principle deliver comparable contributions to the 
DM abundance. In this case, less DM needs to be produced by the individual mechanisms and taking this into 
account is expected to result in a shift of the respective bounds, making the allowed 
mass range slightly larger. The bounds that we provide are thus to be regarded as rather conservative in 
this respect.

%%%%%%%%%%%%%%%%%%%%%%%%%%%%%%%%%%%%%%%%%%%%%%%%%%%%
\subsection{Constraints from self-annihilation}\label{sec:sa}
%%%%%%%%%%%%%%%%%%%%%%%%%%%%%%%%%%%%%%%%%%%%%%%%%%%%

Another constraint on both our scenarios comes from 
self-annihilation processes in DM halos.\footnote{We are 
very grateful to C.~Garcia-Cely for bringing this to our attention.} 
For large enough velocities, two DM particles can annihilate into a pair of
heavier spin-2 modes which in turn may decay into SM fields. Such processes would 
predict signals in direct detection experiments and, if sufficiently strong,
significantly reduce the DM abundance.
We therefore need to make sure that they do not have a relevant impact. 

The self-annihilation process is kinematically forbidden for kinetic energies that
are smaller than the difference in the spin-2 masses, which for $\alpha\ll 1$ is 
given by $\mu_M-\mu_\chi\simeq\frac{1}{2}\alpha^2 m$. 
The specific thermal energy in a halo of DM is estimated by its velocity 
dispersion $\sigma_{\mathrm{DM}}^2$. The biggest clusters have masses 
around $10^{15}\,M_{\odot}$ and a velocity dispersion 
$(\sigma_{\mathrm{DM}}/c)\sim 10^{-3}\,$ (see~e.g.~\cite{Evrard:2007py}).
Hence for $\alpha\gtrsim10^{-3}$ we do not expect self-annihilation to significantly decrease
the DM abundance.\footnote{We have not taken into account particles whose velocities 
reside in the Boltzmann tail of the distribution. Therefore our
bound should be regarded as an order-of-magnitude estimate.}

In fact, this is a very conservative bound on $\alpha$ because 
the two $\delta M_{\mu\nu}$ particles can also
re-annihilate into a $\delta \chi_{\mu\nu}$ pair and 
in certain parameter regions this process may dominate over the decay.
Determining the precise constraints on the parameters for the reverse process 
to be dominant requires making assumptions for the local DM density 
and the velocity distributions in DM halos. We leave these interesting
investigations for future work.

Even if the direct annihilation of $\delta\chi$ into a $\delta M$ pair is kinematically
forbidden, another relevant process could be via an on-shell and a virtual $\delta M$
with a pair of SM particles in the final state.\footnote{We thank an anonymous referee
for bringing up this idea.} For the fastest DM particles with kinetic energy
$\frac{1}{2}\alpha^2 m$ just at the bound, the virtual $\delta M$ 
can be only slightly off-shell such that its propagator contributes with a factor of maximally
$\frac{1}{\alpha \mu_\chi^2}$ to the amplitude. The $\alpha^2$ enhancement counteracts the 
overall suppression by $1/\mpl^6$. It would be interesting to compute the precise rates for this process
and investigate whether it could give rise to observable effects in indirect detection experiments.
In general, observable signals from both processes mentioned here could be produced in the 
parameter regions where the self-annihilation of DM and subsequent decay of 
$\delta M_{\mu\nu}$ take place but are not efficient enough to significantly reduce 
the DM abundance. 

Due to the gravitational coupling, the heaviest spin-2 mode 
decays universally into all SM particles and its decay products (e.g.~photons or neutrinos) 
can be observed in indirect detection experiments. The decay rates are identical to
those in the bimetric case, see~\cite{Babichev:2016bxi}. The corresponding spectra have been 
shown to possess distinct features~\cite{Garcia-Cely:2016pse} and would thus allow
for an identification of the spin-2 parent.

%%%%%%%%%%%%%%%%%%%%%%%%%%%%%%%%%%%%%%%%%%%%%%%%%%%%
\section{Discussion}\label{sec:discussion}
%%%%%%%%%%%%%%%%%%%%%%%%%%%%%%%%%%%%%%%%%%%%%%%%%%%%

We have shown that the maximally symmetric trimetric theory contains a DM candidate that for
$\alpha<1$ can be produced in two different ways. 
A scenario with more than three spin-2 fields essentially gives rise to the same phenomenology.
This follows from the fact that all fluctuations $\delta \chi^{(i)}_{\mu\nu}$ have the same mass
which is parametrically lower than that of $\delta M_{\mu\nu}$, see section~\ref{sec:gen}.
Consequently, the phenomenology of multimetric theory with maximal symmetry is identical 
for any number of different species $\delta \chi^{(i)}_{\mu\nu}$
and going beyond the trimetric case does not give rise to new observable phenomena.

From a theoretical perspective, it is interesting to compare the maximally symmetric 
multimetric theory in terms of mass eigenstates to bimetric theory. 
The latter contains one massless mode $\delta G_{\mu\nu}$ and one massive mode $\delta M_{\mu\nu}$
whose relation to $\delta\gmn$ is of exactly the same form as in the multimetric case, c.f.~(\ref{gitome}).
The expanded multimetric action with $\delta \chi^{(i)}_{\mu\nu}$ formally set to zero,
takes on a very similar form as in bimetric theory. We can thus view multimetric theory as an extension of 
bimetric theory including $(N-1)$ additional massive states $\delta \chi^{(i)}_{\mu\nu}$. It is possible
to freeze out the dynamics of all massive states by taking $\alpha^2=\sum_i\alpha_i^2c_i^2$
and thus all $\alpha_i$ to zero.

The parameter region with small $\alpha$ is well understood in bimetric theory.
From the bimetric results obtained in~\cite{Babichev:2016bxi} we expect that, in the 
parameter region with $\alpha\ll 1$ and $m^2\gg \Lambda$, also trimetric theory resembles GR
to an extremely high precision. New effects in cosmological background solutions are typically 
suppressed by factors of $\alpha^2\frac{\Lambda}{m^2}$ which for heavy spin-2 particles require a large 
value for $\alpha$ in order to be observable. Furthermore, corrections to Newton's law enter through an
exponentially suppressed Yukawa potential. In this case new effects at radius $r$ typically 
appear with a factor $\alpha^2\exp(-m r)$. This shows that for $\alpha\ll 1$, there are basically
no constraints on the spin-2 masses coming from cosmological observations or local tests of gravity.
Moreover, the direct observation of a heavy spin-2 particle 
or its production in colliders is not possible because the couplings to matter are simply too weak. 
Thus for $\alpha<1$, the only possibility for an observable signal (in some parameter regions)
is the one mentioned in section~\ref{sec:sa}: The decay products of the heaviest spin-2 particle 
created through self-annihilation in DM halos may be seen in indirect detection experiments.

Since in both of our proposed scenarios the values for $\alpha$ are not as small as in
the bimetric case, we have neglected the effects of enhanced spin-2 self-interactions. In bimetric
theory they could be invoked to lower the bound on the DM mass~\cite{Chu:2017msm} and it would be
interesting to see whether there exist parameter regions in multimetric theory which allow 
for similar scenarios.

The perturbative treatment 
(which we have implicitly applied in all estimations of bounds on the spin-2 mass) 
is proven to be valid only for energies $E$ of the spin-2 particle that satisfy $E< \alpha\mpl$.
For energies that violate the perturbativity requirement, the theory becomes
strongly coupled (just like QCD is at low energies)
and non-perturbative methods are needed to derive precise results for scattering amplitudes. 
As Figures 3 and 4 indicate, this excludes the possibility to obtain straightforward 
results in a large region of parameter space. 
For instance, for energies $E\sim 10^{16}$\,GeV, our results are completely trustable only if 
$\alpha\gtrsim10^{-3}$. This estimated 
bound is comparable to the one obtained from forbidding all self-annihilation in DM halos
and suggests to pay particular attention to not too small values of $\alpha$.

%%%%%%%%%%%%%%%%%%%%%%%%%%%%%%%%%%%%%%%%%%%%%%%%%%%%%%%%%%%%%%%%
\subsection*{Larger values for $\alpha$}
%%%%%%%%%%%%%%%%%%%%%%%%%%%%%%%%%%%%%%%%%%%%%%%%%%%%%%%%%%%%%%%%

As we stated earlier, a combination of the parameter $\alpha$ and the spin-2 mass scale $m$
controls deviations from GR. 
The existing bounds on spin-2 masses are mostly obtained in the massive gravity limit 
$\alpha\rightarrow\infty$ of bimetric theory where the gravitational 
force mediator is purely massive.
These bounds come from gravitational wave observations, for instance, which give 
$m\lesssim10^{-22}$\,eV \cite{Abbott:2016blz} (for an $\alpha$-dependent bound 
see~\cite{Max:2017flc}). Even stronger bounds are obtained from solar system tests
and weak lensing which require at least $m\lesssim10^{-32}$\,eV~\cite{Dvali:2002vf, deRham:2016nuf}.
Bimetric theory with small spin-2 mass and $\alpha\gg 1$ has also been studied in the context of 
degravitation~\cite{Platscher:2016adw}.

In the bimetric DM scenario of~\cite{Babichev:2016bxi}, 
the constraints from indirect detection forced us to the region with $\alpha \lesssim10^{-12}$.
Multimetric theory, on the other hand, not only allows for spin-2 DM with 
$10^{-12}\lesssim \alpha\lesssim 1$,
but also opens up the possibility for new scenarios in the parameter
region with $\alpha> 1$.
We expect the DM phenomenology to significantly change in this case.
For $\alpha< 1$, the two spin-2 masses in trimetric theory were of the same order of magnitude
but for $\alpha\gg1$ the mode $\delta M_{\mu\nu}$ will become much heavier and decay 
into the lighter spin-2 particle. The analysis of production mechanisms in the previous subsections 
needs to be redone carefully because the vertices in the production diagrams will now be dominated
by contributions with different dependences on $\alpha$.
Moreover, large values for $\alpha$ will result in a new perturbativity bound.
The interesting observation is that 
the lighter spin-2 field $\delta\chi_{\mu\nu}$ may now have a rather low mass since 
$\mu_\chi \simeq \alpha^{-2}\mu_M$ for $\alpha\gg1$.
The presence of a less heavy spin-2 field together with a larger value for $\alpha$ may give rise to 
observational effects in trimetric theory. 
It is therefore an important task for the future to
explore the region $\alpha>1$ and
determine the precise astrophysical and cosmological constraints on the 
lighter spin-2 mass $\mu_\chi$ and the parameter $\alpha^2=\frac{\mu_M^2}{\mu_\chi^2}-1$.

\acknowledgments
We thank T.~Delahaye for useful discussions and 
C.~Garcia-Cely for very valuable comments on the draft of our paper.
This research was partially funded by the bilateral DAAD-CONICYT grant 72150534 (NLGA).
Feynman diagrams were generated with the \textit{TikZ-Feynman} package~\cite{Ellis:2016jkw}.
Some of our calculations have been performed using the \textit{xTensor package} for
Mathematica~\cite{Brizuela:2008ra}.

\appendix
%%%%%%%%%%%%%%%%%%%%%%%%%%%%%%%%%%%%
%%%%%%%%%%%%%%%%%%%%%%%%%%%%%%%%%%%%%%%%%%%%%%%%%%%%%%%%%%%%%%
\section{Quadratic action and cubic vertices}\label{app:cubic}
%%%%%%%%%%%%%%%%%%%%%%%%%%%%%%%%%%%%%%%%%%%%%%%%%%%%%%%%%%%%%%

%%%%%%%%%%%%%%%%%%%%%%%%%%%%%%%%%%%%%%%%%%%%%%%%%%%
\subsection{Useful definitions}
%%%%%%%%%%%%%%%%%%%%%%%%%%%%%%%%%%%%%%%%%%%%%%%%%%%

In order to facilitate the writing down of the action, 
in the following we will simply let $\gmn$ to stand for $\bgmn$ everywhere
and define some useful quantities. First, the bilinear operator,
\begin{align}
K^{(2)}_{\mu\nu}[h,\ell]\equiv&\,
\nabla_\mu h_{\rho\sigma}\nabla_\nu \ell^{\rho\sigma}-\nabla_\mu h\nabla_\nu \ell
+\nabla^\rho h_{\rho\mu}\nabla_\nu\ell+\nabla_\nu h_{\mu\rho}\nabla^\rho\ell
-\nabla_\rho h_{\mu\nu}\nabla^\rho\ell+\nabla_\rho h^{\rho\sigma}\nabla_\sigma\ell_{\mu\nu}\nn\\
&-2\nabla_\mu h^{\rho\sigma}\nabla_\sigma\ell_{\nu\rho}+\nabla_\mu h\nabla^\rho\ell_{\rho\nu}
+\nabla^\rho h_{\mu\nu}\nabla^{\sigma}\ell_{\rho\sigma}
-2\nabla_\rho h_{\mu\sigma}\nabla_\nu\ell^{\rho\sigma}
-2\nabla^\rho h_{\mu\sigma}\nabla^\sigma\ell_{\nu\rho}\nn\\
&+2\nabla^\rho h_{\mu\sigma}\nabla_\rho\ell_\nu^{\ph{\nu}\sigma}
+\nabla^\rho h\nabla_\nu\ell_{\mu\rho}-\nabla^\rho h\nabla_\rho\ell_{\mu\nu}\,.
\end{align}
where $\nabla$ is the covariant derivative compatible with the background metric ${g}_{\mu\nu}$.
Then the combinations,
\beqn
C^{(1)}_{\mu\nu}[h]&\equiv& 2h_{\mu\nu}-\gmn h\,,\\
P^{(1)}_{\mu\nu}[h]&\equiv& h_{\mu\nu}-\gmn h\,.
\eeqn
and,\footnote{We note that these satisfy the relations,
$$C^{(2)}_{\mu\nu}=4h_{\mu}^{\ph{\mu}\rho}C^{(1)}_{\rho\nu}
-\gmn h^{\rho\sigma}C^{(1)}_{\rho\sigma}\,,\quad
P^{(2)}_{\mu\nu}=4h_{\mu}^{\ph{\mu}\rho}P^{(1)}_{\rho\nu}
-\gmn h^{\rho\sigma}P^{(1)}_{\rho\sigma}\,.$$}
\beqn
C^{(2)}_{\mu\nu}[h]&\equiv& 8h_{\mu\rho}h^\rho_{\ph{\rho}\nu}-4hh_{\mu\nu}
-2\gmn h_{\rho\sigma}h^{\rho\sigma}+\gmn h^2\,,
\\
P^{(2)}_{\mu\nu}[h]&\equiv& 4h_{\mu\rho}h^\rho_{\ph{\rho}\nu}-4hh_{\mu\nu}
-\gmn h_{\rho\sigma}h^{\rho\sigma}+\gmn h^2\,.
\eeqn
Finally, we define,
\be
Q^{(2)}_{\mu\nu}[h]\equiv 4h_{\mu\rho}h^\rho_{\ph{\rho}\nu}-3hh_{\mu\nu}
-2\gmn h_{\rho\sigma}h^{\rho\sigma}+\gmn h^2\,.
\ee
We note that in terms of the above operators the GR Lagrangian to cubic order 
in perturbations can be written quite succinctly as,
\begin{align}
\mathcal{L}_{\mathrm{GR}}[h]=\sqrt{-g}\Biggl[
-\frac{1}{12}g^{\mu\nu}K^{(2)}_{\mu\nu}[h,h]
&+\frac{1}{4m_{\mathrm{Pl}}}\left(h^{\mu\nu}-\frac1{6}h\,g^{\mu\nu}\right)K^{(2)}_{\mu\nu}[h,h]\nn\\
&
+2\Lambda\,m_{\mathrm{Pl}}^2
+\frac{\Lambda}{4}h^{\mu\nu}C^{(1)}_{\mu\nu}[h]
-\frac{\Lambda}{12\mpl}h^{\mu\nu}C^{(2)}_{\mu\nu}[h]\Biggr]\,.
\end{align}
Here the first line contains the kinetic operators while the second line 
give all the non-derivative self-interactions up to cubic order.\footnote{With 
an irrelevant $0^{\mathrm{th}}$ order constant which can be removed by adding 
the following non-dynamical term to the action $$
-2m_{\mathrm{Pl}}\Lambda\int\td^4x\sqrt{|g|}\,.$$}

For the maximally symmetric trimetric theory we have,
\be
\beta_n\equiv\alpha_f^{-n}\beta_n^f=\alpha_h^{-n}\beta_n^h\,,\qquad
\beta_1=\beta_3=0\,.
\ee
As explained in section~\ref{sec:cubvert}, we can set,
\be\label{csol}
c_f=\frac{\alpha}{\sqrt{2}\alpha_f}\,,\qquad c_h=\frac{\alpha}{\sqrt{2}\alpha_h}\,.
\ee
without loss of generality. 
Moreover we use~(\ref{lambdadef}) to trade any appearance of $\beta_0$ and $\beta_4$ for 
$\Lambda$ using,
\be\label{betasol}
\beta_0=\frac{1}{2}\left[\frac{\Lambda}{m^2}-3\alpha^2\beta_2\right]\,,\qquad
\beta_4=\frac{2}{\alpha^2}\left[\frac{\Lambda}{m^2}-3\beta_2\right]
\ee 
We will also need the expressions for the spin-2 masses in order to replace occurrences of $\beta_2$,
\be
\mu_\chi^2=2\beta_2m^2\,,\qquad\mu_M^2=2(1+\alpha^2)\beta_2m^2\,.
\ee

%%%%%%%%%%%%%%%%%%%%%%%%%%%%%%%%%%%%%%%%%%%%%%%%%%%%%%%%%%%%%%
\subsection{Trimetric action expanded to cubic order}
%%%%%%%%%%%%%%%%%%%%%%%%%%%%%%%%%%%%%%%%%%%%%%%%%%%%%%%%%%%%%

We are now ready to write down the trimetric action \eqref{triact} expanded 
to cubic order in the mass eigenstates. We write the Lagrangian on the form,
\be
\mathcal{L}_{\mathrm{TM}}=\mathcal{L}_{\mathrm{TM}}^{(0)}
+\mathcal{L}_{\mathrm{TM}}^{(1)}+\mathcal{L}_{\mathrm{TM}}^{(2)}
+\mathcal{L}_{\mathrm{TM}}^{(3)}+\dots\,,
\ee
where the first terms in the expansion are just the trivial ones,
\be
\mathcal{L}_{\mathrm{TM}}^{(0)}=2\Lambda m_{\mathrm{Pl}}^2\sqrt{|g|}\,,\qquad
\mathcal{L}_{\mathrm{TM}}^{(1)}=0\,.
\ee
The first term here can be removed by adding a non-dynamical contribution to the 
action which essentially removes the background spacetime volume integration. 
The second of these terms vanishes due to the background equations. 
The quadratic (or free) Lagrangian terms are given by,
\begin{align}
\frac{\mathcal{L}_{\mathrm{TM}}^{(2)}}{\sqrt{|g|}}=&\,
-\frac{1}{12}g^{\mu\nu}K^{(2)}_{\mu\nu}[\delta G,\delta G]
-\frac{1}{12}g^{\mu\nu}K^{(2)}_{\mu\nu}[\delta M,\delta M]
-\frac{1}{12}g^{\mu\nu}K^{(2)}_{\mu\nu}[\delta \chi,\delta \chi]\nn\\
&+\frac{\Lambda}{4}\delta G^{\mu\nu}C^{(1)}_{\mu\nu}[\delta G]
+\frac{\Lambda}{4}\delta M^{\mu\nu}C^{(1)}_{\mu\nu}[\delta M]
+\frac{\Lambda}{4}\delta\chi^{\mu\nu}C^{(1)}_{\mu\nu}[\delta\chi]\nn\\
&
-\frac{\mu_M^2}{4}\delta M^{\mu\nu}P^{(1)}_{\mu\nu}[\delta M]
-\frac{\mu_\chi^2}{4}\delta\chi^{\mu\nu}P^{(1)}_{\mu\nu}[\delta\chi]\,.
\end{align}
These manifestly correspond to one massless and two massive decoupled spin-2 fields 
propagating on a constant curvature background. The first line contains the canonical Fierz-Pauli 
kinetic terms for spin-2 fields while the second line provides the quadratic response 
to a constantly curved background for such fields, and the third line provides 
the self-interactions giving rise to masses of the fields.
 
The cubic interaction terms are given by,
\begin{align}
\frac{\mathcal{L}_{\mathrm{TM}}^{(3)}m_{\mathrm{Pl}}}{\sqrt{|g|}}&=\,
\frac1{4}\left(\delta G^{\mu\nu}-\frac1{6}\delta Gg^{\mu\nu}\right)\Biggl(
K^{(2)}_{\mu\nu}[\delta G,\delta G]+K^{(2)}_{\mu\nu}[\delta M,\delta M]
+K^{(2)}_{\mu\nu}[\delta \chi,\delta \chi]\Biggr)\nn\\
&
+\frac1{2}\left(\delta M^{\mu\nu}-\frac1{6}\delta Mg^{\mu\nu}\right)\Biggl(
K^{(2)}_{\mu\nu}[\delta G,\delta M]+\frac{1}{2\alpha}K^{(2)}_{\mu\nu}[\delta \chi,\delta \chi]
+\frac{1-\alpha^2}{2\alpha}K^{(2)}_{\mu\nu}[\delta M,\delta M]\Biggr)\nn\\
&
+\frac{1}{2}\left(\delta \chi^{\mu\nu}-\frac1{6}\delta \chi g^{\mu\nu}\right)\Biggl(
K^{(2)}_{\mu\nu}[\delta G,\delta\chi]+\frac{1}{\alpha}K^{(2)}_{\mu\nu}[\delta M,\delta\chi]\Biggr)\nn\\
&
-\frac{\Lambda}{12}\delta G^{\mu\nu}\Biggl(C^{(2)}_{\mu\nu}[\delta G]
+3C^{(2)}_{\mu\nu}[\delta M]+3C^{(2)}_{\mu\nu}[\delta \chi]\Biggr)\nn\\
&
-\frac{\Lambda}{12\alpha}\delta M^{\mu\nu}\Biggl(
(1-\alpha^2)C^{(2)}_{\mu\nu}[\delta M]+3C^{(2)}_{\mu\nu}[\delta\chi]\Biggr)\nn\\
&-\frac{1}{4}\delta G^{\mu\nu}\Biggl(\mu_M^2P^{(2)}_{\mu\nu}[\delta M]
+\mu_\chi^2P^{(2)}_{\mu\nu}[\delta \chi]\Biggr)
-\frac{(1-\alpha^2)}{8\alpha}\mu_M^2\,\delta M^{\mu\nu}P^{(2)}_{\mu\nu}[\delta M]\nn\\
&
+\frac{1}{4\alpha}\delta M^{\mu\nu}\Biggl(
\mu_\chi^2 P^{(2)}_{\mu\nu}[\delta\chi]+\frac{\mu_M^2}{2}Q^{(2)}_{\mu\nu}[\delta\chi]\Biggr)\,.
\end{align}
The first three lines here correspond to the kinetic couplings, 
while the fourth and fifth line come from self-interactions due 
to the background curvature. 
The final two lines arise from interactions due to the mass terms.

%%%%%%%%%%%%%%%%%%%%%%%%%%%%%%%%%%%%%%%%%%%%%%%

 \end{document}